# Brain rhythms in cognition – controversies and future directions

*Anne Keitel\*, Christian Keitel\*, Mohsen Alavash, Karin Bakardjian, Christopher S.Y. Benwell, Sophie Bouton, Niko A. Busch, Antonio Criscuolo, Keith B. Doelling, Laura Dugué, Laetitia Grabot, Joachim Gross, Simon Hanslmayr, Laura-Isabelle Klatt, Daniel S. Kluger, Gemma Learmonth, Raquel E. London, Christina Lubinus, Andrea E.  Martin, Jonas Obleser, Johanna M. Rimmele, Vincenzo Romei, Manuela Ruzzoli, Felix Siebenhühner, Sophie Slaats, Eelke Spaak, Luca Tarasi, Gregor Thut, Jelena Trajkovic, Danying Wang, Malte Wöstmann, Benedikt Zoefel, Satu Palva⁺, Paul Sauseng⁺, Sonja A. Kotz⁺*

\* Joint first authorship
⁺ Joint senior authorship

The remaining authors are listed in alphabetical order

List of authors and affiliations

| Name | Affiliation | ⓘ ORCID |
|---|---|---|
| Anne Keitel | Psychology Division, School of Humanities, Social Sciences and Law, University of Dundee, Dundee, UK | 0000-0003-4498-0146 |
| Christian Keitel | Psychology Division, School of Humanities, Social Sciences and Law, University of Dundee, Dundee, UK | 0000-0003-2597-5499 |
| Mohsen Alavash | Department of Psychology, University of Lübeck, Lübeck, Germany | 0000-0002-4166-2321 |
| Karin Bakardjian | Faculty of Psychology and Neuroscience, Maastricht University, Maastricht, The Netherlands. | |
| Christopher S.Y. Benwell | Psychology Division, School of Humanities, Social Sciences and Law, University of Dundee, Dundee, UK | 0000-0002-4157-4049 |
| Sophie Bouton | Université Paris Cité, Institut Pasteur, AP-HP, Inserm, CNRS, Fondation Pour l'Audition, Institut de l'Audition, IHU reConnect, Paris, France | 0000-0001-5496-4583 |
| Niko A. Busch | Institute of Psychology, University of Münster, Münster, Germany | 0000-0003-4837-0345 |
| Antonio Criscuolo | Department of Neuropsychology & Psychopharmacology, Faculty of Psychology and Neuroscience, Maastricht University | 0000-0003-3192-441X |

| Keith B. Doelling | Université Paris Cité, Institut Pasteur, AP-HP, INSERM, CNRS, Fondation Pour l'Audition, Institut de l'Audition, IHU reConnect, F-75012 Paris, France | 0000-0002-0776-8114 |
|---|---|---|
| Laura Dugué | Université Paris Cité, CNRS, Integrative Neuroscience and Cognition Center, Paris, France; Institut Universitaire de France (IUF), Paris, France | 0000-0003-3085-1458 |
| Laetitia Grabot | Laboratoire des Systèmes Perceptifs (LSP), Département d'études cognitives, École Normale Supérieure, PSL University, CNRS, Paris, France | 0000-0001-9987-5883 |
| Joachim Gross | Institute for Biomagnetism and Biosignal Analysis, University of Münster, Münster, Germany | 0000-0002-3994-1006 |
| Simon Hanslmayr | School of Psychology and Neuroscience, University of Glasgow, UK | 0000-0003-4448-2147 |
| Laura-Isabelle Klatt | Leibniz Research Centre for Working Environment and Human Factors, Dortmund, Germany | 0000-0002-5682-5824 |
| Daniel S. Kluger | Institute for Biomagnetism and Biosignal Analysis, University of Münster, Münster, Germany | 0000-0002-0691-794X |
| Gemma Learmonth | Division of Psychology, University of Stirling, UK | 0000-0003-4061-4464 |
| Raquel E. London | Department of Experimental Psychology, Faculty of Psychology and Educational Sciences, Ghent University, Ghent, Belgium | 0000-0003-1678-2556 |
| Christina Lubinus | Department of Cognitive Neuropsychology, Max-Planck-Institute for Empirical Aesthetics, Frankfurt am Main, Germany | 0000-0002-7261-2490 |
| Andrea E. Martin | Language and Computation in Neural Systems Group, Max Planck Institute for Psycholinguistics & Donders Centre for Cognitive Neuroimaging, Radboud University, Nijmegen, The Netherlands | 0000-0002-3395-7234 |
| Jonas Obleser | Department of Psychology, University of Lübeck, Lübeck, Germany | 0000-0002-7619-0459 |
| Johanna M. Rimmele | Department of Cognitive Neuropsychology, Max-Planck-Institute for Empirical Aesthetics, Frankfurt am Main, Germany | 0000-0002-2065-9772 |

| Vincenzo Romei | Centro studi e ricerche in Neuroscienze Cognitive, Dipartimento di Psicologia, Alma Mater Studiorum - Università di Bologna, Campus di Cesena, Cesena, Italy; Facultad de Lenguas y Educación, Universidad Antonio de Nebrija, Madrid, Spain | 0000-0003-1214-2316 |
|---|---|---|
| Manuela Ruzzoli | Basque Center on Cognition Brain and Language (BCBL), Donostia/San Sebastián, Spain; Ikerbasque, Basque Foundation for Science, Bilbao, Spain | 0000-0002-1719-7140 |
| Felix Siebenhühner | Neuroscience Center, Helsinki Institute of Life Science, University of Helsinki, Helsinki, Finland; Department of Neuroscience and Bioengineering (NBE), Aalto University, Espoo, Finland | 0000-0002-2621-9145 |
| Sophie Slaats | Department of Basic Neurosciences, Faculty of Medicine, University of Geneva, Geneva, Switzerland | 0000-0001-9596-9073 |
| Eelke Spaak | Donders Institute, Radboud University, Nijmegen, The Netherlands | 0000-0002-2018-3364 |
| Luca Tarasi | Centro Studi e Ricerche in Neuroscienze Cognitive, Dipartimento di Psicologia, Alma Mater Studiorum – Università di Bologna, Campus di Cesena, Cesena, Italy | 0000-0002-4977-3173 |
| Gregor Thut | Centre de Recherche Cerveau et Cognition (Cerco), CNRS UMR5549 and Université de Toulouse, Toulouse, France | 0000-0003-1313-4262 |
| Jelena Trajkovic | Department of Cognitive Neuroscience, Faculty of Psychology and Neuroscience, Maastricht University, Netherlands | 0000-0002-1215-2410 |
| Danying Wang | Department of Neuroscience, Physiology and Pharmacology, University College London, UK; School for Psychology and Neuroscience and Centre for Cognitive Neuroimaging, University of Glasgow, UK | 0000-0003-0543-7036 |
| Malte Wöstmann | Department of Psychology, University of Lübeck, Lübeck, Germany | 0000-0001-8612-8205 |
| Benedikt Zoefel | Université de Toulouse, CNRS, Centre de Recherche Cerveau et Cognition (CerCo), Toulouse, France | 0000-0002-9800-2551 |
| Satu Palva | Neuroscience Center, Helsinki Institute of Life Science, University of Helsinki, Finland; School of Psychology and Neuroscience, University of Glasgow, UK | 0000-0001-9496-7391 |
| Paul Sauseng | Department of Psychology, University of Zürich, Zürich, Switzerland | 0000-0003-3410-3895 |

| Sonja A. Kotz | Dept. of Neuropsychology and Psychopharmacology, Faculty of Psychology and Neuroscience, Maastricht University, Maastricht, The Netherlands; Dept. of Neuropsychology, Max Planck Institute for Human Cognitive and Brain Sciences, Leipzig, Germany | 0000-0002-5894-4624 |

## Abstract

Brain rhythms seem central to understanding the neurophysiological basis of human cognition. Yet, despite significant advances, key questions remain unresolved. In this comprehensive position paper, we review the current state of the art on oscillatory mechanisms and their cognitive relevance. The paper critically examines physiological underpinnings, from phase-related dynamics like cyclic excitability, to amplitude-based phenomena, such as gating by inhibition, and their interactions, such as phase-amplitude coupling, as well as frequency dynamics, like sampling mechanisms. We also critically evaluate future research directions, including travelling waves and brain-body interactions. We then provide an in-depth analysis of the role of brain rhythms across cognitive domains, including perception, attention, memory, and communication, emphasising ongoing debates and open questions in each area. By summarising current theories and highlighting gaps, this position paper offers a roadmap for future research, aimed at facilitating a unified framework of rhythmic brain function underlying cognition.

## Introduction

Brain rhythms - neuronal oscillations - are periodic changes of excitability in neuronal populations. Ubiquitous in the brain, rhythmic activity can be observed at different levels of the neuronal hierarchy as well as at different timescales. We typically consider activity with frequencies between 1-100 Hz, but rhythms can range further into infra-slow or very-high frequency realms. The scientific interest in brain rhythms has soared in recent years, and multidisciplinary efforts have led to considerable progress on putative mechanisms and their importance for perception and cognition. However, burgeoning knowledge and methodological advances have also generated a multitude of new questions.

The main purpose of this extensive consensus paper is to highlight the state-of-the-art knowledge and currently unsolved questions, controversies, and debates related to oscillatory cognitive neuroscience. First, we present our knowledge of what brain rhythms do, i.e., the current idea of fundamental oscillatory mechanisms and functions (**Figure 1**). We also cover oscillatory activity during rest and interactions with other bodily rhythms, a promising sub-field of oscillatory neuroscience. We then provide a snapshot of the role of oscillatory mechanisms in cognitive processes spanning perception, attention, memory and communication, and highlight open questions in each area.

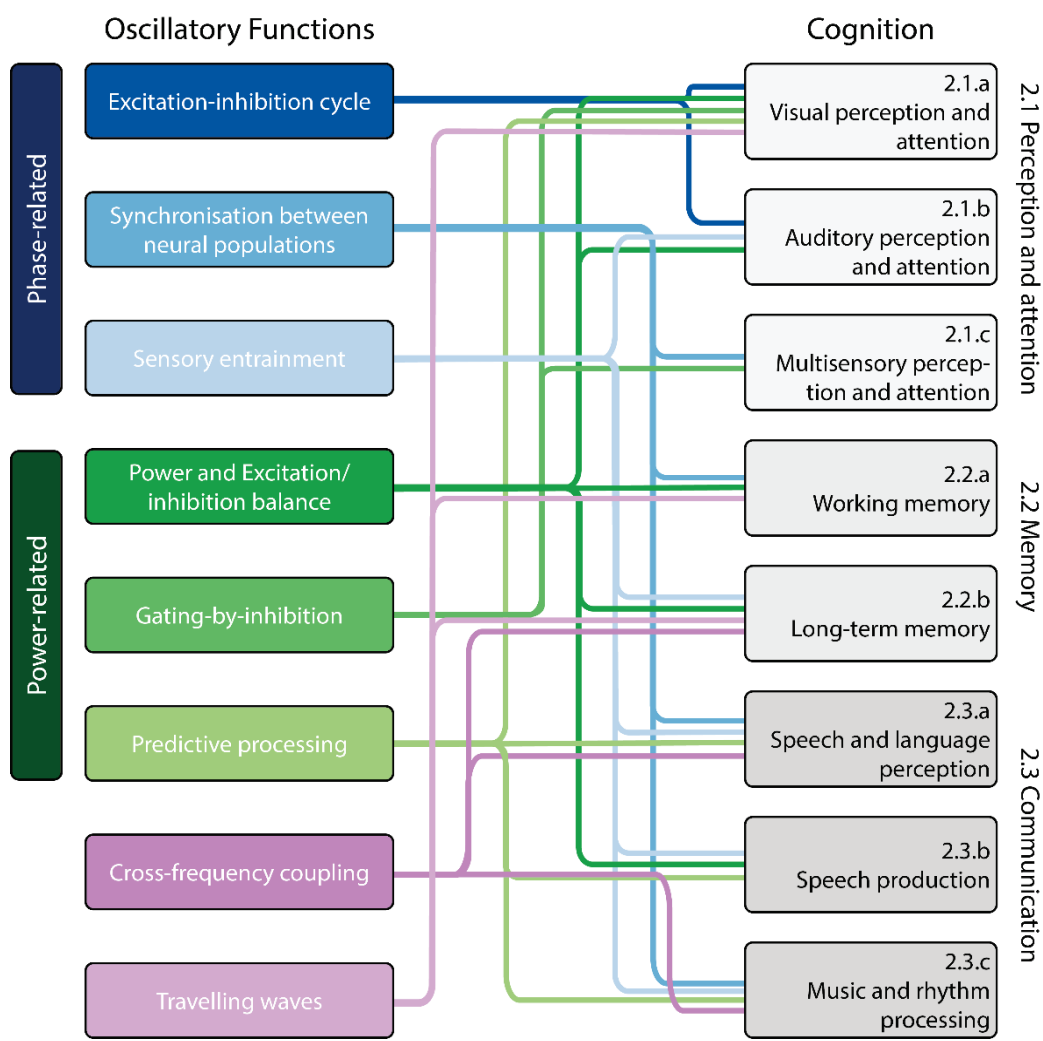

**Figure 1.** Overview of putative oscillatory mechanisms and functions, and cognitive processes in which they are described to be relevant. Note that the conceptual level of proposed functions ranges from close

to neurophysiological processes (e.g., cross-frequency coupling) to cognitive interpretations (e.g., gating by inhibition).

We adopt the widespread approach that observations about rhythmic brain activity and processes involving brain rhythms can largely be described by using concepts from oscillatory mechanics (Buzsáki, 2006). From this perspective, rhythms are quantified as sinusoidal oscillations that undergo changes in power (amplitude), phase (phase shifts, phase resets), and frequency. Consequently, oscillations are typically extracted from electro- or magneto-encephalographical (M/EEG) time-series data by filtering them into narrow frequency bands or using Fourier-based spectral and spectro-temporal decompositions, including fast Fourier transform (FFT) and wavelet convolution (e.g., Gross, 2014).

While this Fourier-based oscillatory view of brain rhythms is convenient, we note that other methodological approaches are available for capturing existing non-sinusoidal or non-stationary characteristics of rhythmic activity (waveform shape: Cole & Voytek, 2019; empirical mode decomposition: Huang *et al.*, 1998; multi-scale entropy: Kosciessa *et al.*, 2020; Myrov *et al.*, 2024) (and burst detection: Hughes *et al.*, 2012; Myrov *et al.*, 2024; Wilson *et al.*, 2022) and that equating the methodological approach with physiological and cognitive interpretations can produce misinterpretations, such as the Fourier fallacy (Jasper, 1948).

# 1    Putative oscillatory mechanisms

Oscillations can be described by their phase, amplitude (or power), and period (usually quantified by frequency, **Figure 2**). These three parameters have been linked to partially distinct physiological mechanisms, functions, and processes. However, defining what constitutes an oscillatory mechanism is far from trivial, and it is often difficult to disentangle different levels of processing, from basic physiological processes at the level of neurons to cognitive processes. Oscillations have intrinsic timescales that depend on the period of the oscillations (cycle). Whereas higher-frequency oscillations occur on shorter time scales, lower-frequency oscillations describe longer cycles.

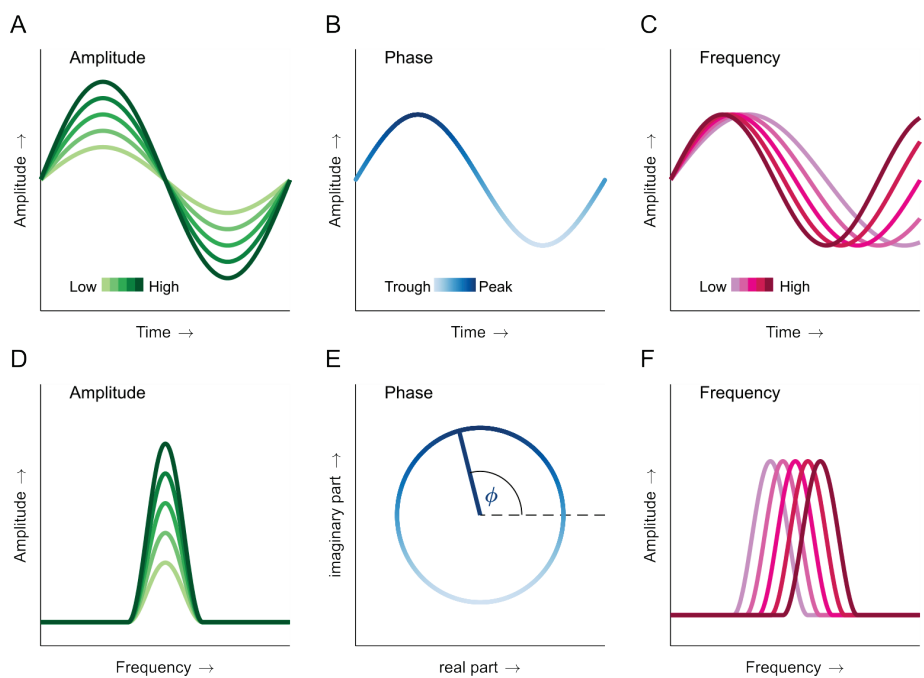

**Figure 2.** Basic properties of oscillations. A-C show one cycle of an oscillation with variations in amplitude, instantaneous phase, and frequency in the time domain. D-F translate these into the spectral domain. Amplitude in D is usually expressed as oscillatory power (= amplitude squared). The phase angle $\phi$ in E expresses the lie of peaks and troughs relative to an oscillation $\phi = 0$. (Note that $\phi$ in E does not match the oscillation in B for illustrative purposes).

At the neuronal level, oscillations reflect the waxing and waning of neuronal excitability, thus regulating neuronal spiking and communication in neuronal circuits. Both neurophysiological evidence and computational biophysical models of neuronal circuit mechanisms suggest that synaptic interactions among excitatory pyramidal neurons (PNs) and GABAergic inhibitory interneurons (INs) form the simplest universal microcircuit that intrinsically generates synchronised oscillations through recurrent and reciprocal interactions (Onslow *et al.*, 2014; Traub *et al.*, 2005; Voloh & Womelsdorf, 2016). The frequency of an oscillation is determined by several factors including the strength of the excitatory drive, axonal conduction delays, and the time constants of GABA-A- and GABA-B-receptor-mediated inhibitory postsynaptic potentials (Buzsáki & Wang, 2012).

Human brain rhythms tend to occur within characteristic frequency bands (Groppe *et al.*, 2013; Mahjoory *et al.*, 2020). Conventionally, oscillations have been separated into canonical bands of lower frequencies in delta (1-4 Hz), theta (4-8 Hz), and alpha (8-14 Hz) ranges, as well as higher frequencies in beta (14-30 Hz) and gamma (30-120 Hz) ranges. These bands are framed by infra-slow (<1Hz) and high-frequency oscillations (HFOs; >120Hz). This classification was originally borne out of the low sensitivity of early EEG recordings and analysis approaches, that simplified a more complex reality of multiple rhythms in overlapping frequency bands that are present in brain activity at any point in time. Recent data-driven approaches show that frequency bands can vary considerably between individuals and brain anatomy (e.g., Haegens *et al.*, 2014) and tasks (e.g., Cruz *et al.*, 2025; Sattelberger *et al.*, 2024) while also demonstrating more fine-grained frequency distributions or sub-bands.

## 1.1 Mechanisms of oscillatory phase

Instantaneous phase describes the momentary state of an oscillating system within its cycle (**Figure 2**). During each cycle, the system traverses different states, including one peak (positive maximum) and one trough (negative maximum). Two oscillators are considered synchronised when their peaks and troughs align or show statistically significant coupling over time or across trials. In this section we discuss what the different phases and phase synchronisation of a neuronal oscillation might reflect, providing the basis for how they can be mechanistically linked to cognitive function (Section 2).

### 1.1.a Excitation-inhibition cycle in oscillation dynamics

Evidence that rhythmicity plays a fundamental role in neuronal processing first arose in the early 20[th] century, when Bishop (1933) applied periodic stimuli to the optic nerve of a rabbit and observed that the amplitude of the cortical responses varied rhythmically. This was taken as evidence that the phase of cortical oscillations reflected the cortex's momentary excitability (Lindsley, 1952). In line with this, later behavioural studies showed that oscillatory phase predicts the sensitivity to sensory stimulation (e.g., visual stimulation: (Busch *et al.*, 2009; Dustman & Beck, 1965; Harris *et al.*, 2018; Lansing, 1957), auditory stimulation: (Kayser *et al.*, 2016; Neuling *et al.*, 2012; Ng *et al.*, 2012). Cycling through phases has therefore been proposed to provide "windows of opportunity" for perceptual processing, also known as "perceptual cycling" (Lindsley, 1952; VanRullen, 2016), a notion revisited in more detail in Section 2.1.

Potentially underpinning this notion at the neuronal level, rhythmic activity plays a fundamental role in the top-down regulation of neuronal processing. High-excitability phases are associated with, or defined by, higher neuronal firing rates compared with low-excitability phases. This structure creates a temporal scaffolding for information processing across frequency bands (Buzsáki, 2006). The same principle is thought to apply to oscillations across multiple frequencies. For example, beta and gamma-band oscillations are associated with tight LFP-spike time relationships and therefore thought to reflect periods of active neuronal processing (Fries, 2015; Womelsdorf *et al.*, 2007). Slower rhythms (1-14 Hz) have been ascribed a modulatory influence by setting the temporal scaffolding for neural processing (Gips *et al.*, 2016; J. M. Palva & S. Palva, 2018; von Stein & Sarnthein, 2000).

In human non-invasive recordings, it is impossible to measure excitability, neuronal firing rates or high-/low-excitability phases in absolute terms. Instead, various indirect measures yield a 'proxy' of these phenomena. For example, phase opposition can be used to distinguish high- and low-excitability phases (VanRullen, 2016). Both invasive and non-invasive recordings have found evidence of distinct high-/low-excitability phases in slow oscillations (< 1 Hz, Bergmann *et al.*, 2012; Vanhatalo *et al.*, 2004), mu oscillations (Zrenner *et al.*, 2023), alpha oscillations (Bishop, 1933; Bollimunta *et al.*, 2008; Haegens *et al.*, 2011), beta oscillations (Mäki & Ilmoniemi, 2010; van Elswijk *et al.*, 2010), and gamma oscillations (Berger *et al.*, 2014).

### 1.1.b  Synchronisation between neuronal populations

Neuronal oscillations are invariably associated with neuronal phase synchronisation (also: phase coherence) between distinct oscillating populations. The general idea pointed out in section 1.1.a, that the phase of an oscillation reflects fluctuations of excitability/inhibition, suggests that a large portion of neurons influenced by the excitation/inhibition cycle should therefore be activated and deactivated synchronously. At the neuronal level, this should be reflected as synchronous neuronal spiking across simultaneously oscillating neuronal assemblies (Fries, 2015). Therefore, neuronal synchronisation has been suggested to underlie the regulation of neuronal processing at the circuit level across spatially distributed neuronal assemblies (Fries, 2015; Singer, 1999). This 'binding-by-synchrony' hypothesis posits that temporally coincident spikes evoke action potentials in downstream neurons more effectively than asynchronous inputs (Singer, 1999; Singer & Gray, 1995), a position that has been challenged subsequently (Ray & Maunsell, 2010; Roelfsema, 2023).

Taking this idea further, the communication-through-coherence (CTC) hypothesis holds that synchronised firing of distinct neuronal populations allows flexible establishment of stable communication channels across the brain (Fries, 2015). More specifically, any two neuronal populations will exchange information most effectively when their firing patterns are aligned through phase coherence of (high-frequency) rhythms. The stability of CTC is grounded in the temporal predictability of upcoming cycles when sender and receiver activity is phase-coherent. Within limits, even the direction of information transfer can be estimated using Granger causality approaches or directed-coherence measures (Bastos *et al.*, 2015b; Bressler *et al.*, 2021; Kaminski *et al.*, 2016).

However, it is well established that GABA-ergic interneurons play a crucial role in generating brain oscillations (Buzsáki, 2006; Fries *et al.*, 2007; Mann & Paulsen, 2007) and that the number of these inhibitory interneurons is far smaller than that of excitatory pyramidal neurons. Thus, phase-coherent oscillations recorded on a macroscale could potentially also arise without any related synchronised spiking of excitatory neurons (Pesaran *et al.*, 2018; Schneider *et al.*, 2021).

In line with this, Schneider *et al.* (2021) recently introduced an alternative to CTC. This account posits that (phase) coherence between sending and receiving populations may merely be a consequence of the transmission. Their "Synaptic-Source-Mixing" (SSM) model therefore challenges the functional role of rhythmic activity in communication between neuronal populations and identifies alternative determinants that do not require oscillatory coupling between sender and receiver: their afferent connectivity, their individual spectral profiles, and the coherence between the sender's local-field potential and the projected signal ('source-projection coherence') itself. According to SSM, variations in inter-areal coherence can largely be explained as a function of afferent connectivity and sender spectral power (see section 1.2.a). Frequency-specific coherence can then arise as a by-product of a strong oscillatory component in the sender projection. Future work will tell whether SSM generally provides a refined perspective that goes beyond CTC, withstanding experimental challenges in which 'causality' is probed by perturbing neuronal populations (Uran *et al.*, 2022) and, more generally, explaining aspects of human cognition. If so, the previously held view of the role of rhythmic activity in 'brain-wide broadcasting' may have to be reconsidered (also see Vinck *et al.*, 2023).

This recent idea notwithstanding, changes in information processing and exchange are currently associated with time-varying phase synchronisation across brain areas (Palva & Palva, 2012; Womelsdorf *et al.*, 2014). These changes may follow modulations in amplitude (see section 1.2) in at least one of the synchronised neuronal populations (Palva & Palva, 2012) or arise as result of discontinuities in instantaneous phase, known as a 'phase reset' (Voloh & Womelsdorf, 2016). However, macroscopic EEG and MEG recordings alone make it difficult to assess whether phase has truly reset or whether a previously silent population of neurons has started generating new rhythmic activity (Sauseng *et al.*, 2007). Including behavioural readouts is one way to better understand the dynamics of phase synchronisation. This makes it easier to identify synchronised activity that reflects actual communication between brain areas (Palva & Palva, 2012).

Experimentally, synchronisation as a means of communication between neuronal populations in distinct brain areas has also been probed with rhythmic neurostimulation. Studies using bifocal transcranial magnetic stimulation (TMS) or transcranial alternating current stimulation (tACS), i.e., externally applied magnetic or electrical signals that influence endogenous brain rhythms, suggest that phase synchronisation between distant regions can be manipulated (Biel *et al.*, 2022; Plewnia *et al.*, 2008; Polania *et al.*, 2012; Salamanca-Giron *et al.*, 2021). Critically, this approach emulates an oscillatory process – entrainment – that is thought to be involved in establishing naturally occurring phase synchronisation between neuronal populations.

Entrainment occurs when a rhythm-generating neuronal population, a 'neural oscillator', synchronises to the input rhythm of a second neuronal population that can be different from the natural, or eigen-frequency of the entrained population (Herrmann *et al.*, 2016). Although sometimes used synonymously, entrainment needs to be distinguished from the effect of resonance, where an external rhythm drives an increase in the amplitude of a neural oscillator at its eigenfrequency. Both phenomena depend on the strength of the external drive and the difference of driving frequency and neural eigenfrequency. Resonance can transition into entrainment when the input drive is strong enough. Note that, while Communication-Through-Coherence requires entrainment, newer frameworks, also based on the idea of Synaptic Source Mixing mentioned above, challenge its purported role in establishing synchrony-based neuronal communication (Vinck *et al.*, 2023).

### 1.1.c Sensory entrainment

Sensory entrainment is a special case of entrainment that assumes that some rhythm-generating neuronal populations, i.e., self-sustained neural oscillators, can also synchronise to periodicities in external sensory input (Lakatos *et al.*, 2019). Although widely assumed, genuine sensory entrainment has been challenging to observe in neurophysiological data, due to methodological limitations in separating its spectral signatures from those of stimulus-evoked responses driven by periodic input signals (Duecker *et al.*, 2024; Keitel *et al.*, 2019; Zoefel *et al.*, 2018b). Put differently, it is not trivial to discern whether a spectral peak at the stimulation frequency, as observed in neurophysiological recordings following periodic stimulation, reflects the tracking of periodic input dynamics (Capilla *et al.*, 2011; Keitel *et al.*, 2017) or indicates the involvement of entrained neural oscillators (Doelling *et al.*, 2019; Gulbinaite *et al.*, 2019). Realistically, it is reasonable to assume that periodic stimulation, as any other type of stimulation, will always produce a stimulus-evoked response. From this perspective, sensory entrainment should therefore manifest as indications of entrained neural oscillators *in addition* to stimulus-related processing in measures of stimulus-brain coupling.

Direct or even indirect evidence for sensory entrainment, for example in the context of auditory stimulation, such as continuous speech and music, is sparse (but see, Kösem *et al.*, 2018; Lakatos *et al.*, 2013; van Bree *et al.*, 2021). Moreover, most naturalistic stimuli contain multiple, hierarchical stimulation frequencies (Garcia-Rosales *et al.*, 2018). Therefore, if entrainment occurs in addition to stimulus-evoked responses, another question is whether it does so on all levels of the hierarchical input, or only on a select few levels. For example, in the case of speech, entrainment might only occur at the relatively 'rhythmic' syllable rate typically found in the theta frequency range, but not at the word or phoneme rates (see section 2.3.a).

In sum, sensory entrainment is an often-assumed mechanism for the processing of (quasi-) rhythmic input, but it is still unclear under which circumstances stimulus-brain coupling goes beyond evoked neural responses. Researchers have adopted refined terminologies, such as distinguishing between cortical 'tracking' vs. 'entrainment', or 'entrainment in a broad sense' (Obleser & Kayser, 2019) when brain-stimulus coupling is observed, but the involvement of endogenous neural oscillators has not been established. Although these distinctions seem semantic, they remain relevant for the correct interpretation of empirical findings and may be consequential for applications that use rhythmic sensory stimulation such as Brain-Computer-Interfaces (Liu *et al.*, 2022) or potentially effective interventions in neurodegenerative diseases (but see, He *et al.*, 2021; Iaccarino *et al.*, 2016; Yang & Lai, 2023).

## 1.2 Mechanisms of oscillatory power

Oscillatory amplitude or power are measures of the strength of oscillations (**Figure 2**). Similar to phase, amplitude can be interpreted more meaningfully if oscillations are observed over several cycles, indicating the presence of a genuinely rhythmic activity. In this section, we introduce the neural processes that are hypothesised to underlie neuronal power dynamics, laying the groundwork for understanding their role in cognitive function (see section 2).

### 1.2.a Macroscopic measures of power and excitation-inhibition balance in oscillation dynamics

Phase and amplitude are intertwined features of oscillations – the more neurons are in phase, the higher the amplitude in a specific frequency band will be. Higher amplitude is therefore taken to reflect more wide-spread synchronisation within a neural population. However, amplitude modulations are also often discussed in terms of inhibition such that amplitude/power levels of low frequencies from theta to low beta bands are considered to reflect a state of inhibition (e.g., Zarkowski *et al.*, 2006), specifically in the alpha band (Jensen & Mazaheri, 2010; Klimesch *et al.*, 2007). This idea stems from findings that show that alpha oscillations are associated with decreased neuronal firing rates (Bollimunta *et al.*, 2008; Haegens *et al.*, 2011). This idea is further supported by findings that TMS-induced phenomena, such as visual phosphenes and myographic activity, are decreased with stronger alpha in respective visual or motor cortices (Romei *et al.*, 2008a; Samaha *et al.*, 2017a; Sauseng *et al.*, 2009a; Zarkowski *et al.*, 2006). However, this view is also debated as not all evidence supports with this framework (Palva & Palva, 2007, 2011).

A further line of research investigates how excitation/inhibition (E/I) balance, i.e. the ratio between excitatory and inhibitory signals, affects the power spectrum. Different approximations have been developed to infer the underlying E/I balance from non-invasive EEG / MEG data. These approaches consider the functional E/I balance (Bruining *et al.*, 2020) or slope of the 1/f power spectrum (He, 2014; Plenz & Thiagarajan, 2007). A flatter slope (i.e., relatively lower power at low frequencies and higher power at high frequencies) has been proposed to indicate a state of excitation, while a steeper slope indicates a state of inhibition (Ahmad *et al.*, 2022; Gao *et al.*, 2017). In parallel, the 1/*f*-like shape of the power spectrum and the spatiotemporally scale-free nature of brain dynamics have been interpreted as evidence that the brain operates near criticality (He, 2014; O'Byrne & Jerbi, 2022; S. Palva & J. M. Palva, 2018; Plenz & Thiagarajan, 2007), with these dynamics and aperiodic activity (Gyurkovics *et al.*, 2022; Preston *et al.*, 2025) playing fundamental roles in neuronal processing.

Additionally, the relationship between oscillatory rhythms and blood oxygenation (BOLD), measured with functional magnetic resonance imaging (fMRI), has been frequently studied. Most of these studies have found negative correlations between alpha activity and the BOLD signal, indicating that strong alpha amplitude is associated with reduced BOLD signal in occipital and parietal cortex (Goldman *et al.*, 2002; Laufs *et al.*, 2003; Moosmann *et al.*, 2003), but positive correlations have also been found, for example in the thalamus, insula, and cingulate cortex (Goldman *et al.*, 2002; Moosmann *et al.*, 2003; Sadaghiani *et al.*, 2010). More recently, ultra-high field 7T MR spectroscopy has been used to delineate biological underpinnings of the measure of E/I-balance through 1/f slopes, revealing a contribution of inhibitory GABA activity (McKeon *et al.*, 2024). In summary, there is a complex relationship between oscillation amplitudes, 1/f confounds, E/I balance, and metabolic activity, which is reflected in the complexity of behavioural functions associated with oscillatory rhythms.

A further issue that complicates the use of measures of oscillatory power is that it is often an underlying assumption that neural activity reflects sustained oscillations. However, it has been shown that some patterns of neural activity are consistent with transient bursts of power, especially in beta and gamma frequency bands (Jones, 2016; Lundqvist *et al.*, 2016). To distinguish between these two alternatives, using single-trial analyses and avoiding trial-wise averaging has been recommended (Lundqvist *et al.*, 2016; Stokes & Spaak, 2016).

### 1.2.b   Gating by inhibition

Alpha amplitude has also been linked to the gating of information of neural signals (Jensen & Mazaheri, 2010), which, in turn, may relate to gain control of neural processes. Gating could occur at the network level affecting the transfer of information between nodes. This is in line with a view that alpha-amplitude (and -phase) dynamics reflect a network phenomenon (Palva &

Palva, 2007). Alternatively, gating could occur at a local level in line with the notion that alpha-amplitude fluctuations reflect up- or down-regulation of local excitability (Romei *et al.*, 2008a). These two mechanisms are not mutually exclusive. There is evidence that alpha influences both, network and local activity, where a higher-lower band dissociation reflects network versus local effects (Lobier *et al.*, 2018). Orthogonal to this view are considerations on the level at which alpha amplitude may gate processing of information (early versus late, or input versus access of information). As alpha is a thalamocortical rhythm including geniculo-cortical connections (Lorincz *et al.*, 2009), gating may occur at an early input stage of sensory information processing. Gating at this level may relate to gain modulation of early responses in V1, such as modulation of early visual evoked potentials (Trajkovic *et al.*, 2024); but see, (Morrow *et al.*, 2023). On the other hand, alpha oscillations are also observed in pulvino-cortical networks associated with higher-order cognition where they have been shown to regulate synchronisation between cortical areas (Saalmann *et al.*, 2012). The latter findings highlight alpha's role in cortico-cortical interactions. This may point towards a stronger role in the gating of information at later processing stages in the visual stream rather than early geniculo-cortical input.

### 1.2.c   Predictive processing

An influential idea in cognitive neuroscience is that of the 'Bayesian brain': the brain is continuously trying to figure out the hidden causes of its often-noisy input (Clark, 2013; Friston, 2010; Rao & Ballard, 1999). Computationally, this is well-implemented by the ongoing interaction among prior expectations, the predictions generated from these, the incoming stimuli, and the prediction error resulting from the comparison between those. We refer to the general framework giving centre stage to priors and predictions as 'predictive processing', while a specific influential model, allocating a key role to predictions errors, is known as 'predictive coding' (Bastos *et al.*, 2012).

Neural oscillations in different frequency bands show distinct associations with feedback and feedforward processing, and it is this profile that has led several authors to propose an intimate link between oscillatory activity and predictive processing. Briefly, a commonly held model is that lower-frequency (alpha/beta) activity tends to carry predictions in feedback directions, while higher-frequency (gamma) activity is associated with prediction errors carried forward (Bastos *et al.*, 2012). This is in line with earlier reports that synchrony between frontal and parietal cortex is stronger specifically in low frequencies during top-down driven attention, while it is stronger in high frequencies during bottom-up driven attention (Buschman & Miller, 2007). Recent studies suggest that this communication between brain areas might be enabled by travelling waves (Mohanta *et al.*; Tarasi *et al.*, 2025) (see 1.4. below). The overarching view linking oscillations to feedback/feedforward predictive processing has found support from non-human primate studies (Bastos *et al.*, 2015a), including those with laminar specificity (van Kerkoerle *et al.*, 2014), as well as human MEG (Michalareas *et al.*, 2016).

These lines of evidence notwithstanding, recent work suggests that this elegant mapping of predictions onto low frequencies and feedback, and prediction errors onto high frequencies and feedforward projections, may be overly simplistic (Vinck *et al.*, 2022). For example, highly predictable visual input appears to be associated specifically with high-frequency synchronisation in the visual cortex (Uran *et al.*, 2022). In conclusion, even if the one-to-one mapping between oscillatory frequency bands and predictions or prediction errors has been questioned, it appears likely that neural oscillations are closely associated with predictive processing.

## 1.3   Cross-frequency coupling mechanisms

Cross-frequency coupling (CFC) has been suggested as a putative mechanism that integrates and coordinates different elementary operations involved in cognitive function. Importantly, CFC involves both local (within one population, or within one broadband time series in recorded data) and inter-areal (between populations) mechanisms (Siebenhühner *et al.*, 2020). Established forms of CFC are phase-amplitude, phase-phase, and amplitude-amplitude coupling, although others have also been proposed (Hyafil *et al.*, 2015b; J. M. Palva & S. Palva, 2018).

Phase-amplitude couplings or 'nested oscillations' are the most investigated form of CFC, e. g., in the rodent hippocampus, non-human primates, and in human M/EEG and intracranial EEG. Phase-amplitude coupling reflects the amplitude modulation of a faster oscillation through excitability fluctuations imposed by the phase of lower-frequency oscillations (Canolty & Knight, 2010; Fell & Axmacher, 2011; Lisman & Jensen, 2013; Schroeder & Lakatos, 2009; Spaak *et al.*, 2012). Notably, the strength, frequency ratio, as well as the individual frequencies of PAC can change to adapt to task demands (Canolty & Knight, 2010).

In contrast to phase-amplitude coupling, phase-phase-coupling, or cross-frequency phase synchrony, describes a consistent spike-time relationship between two oscillations similarly to phase synchrony, but at two different frequencies f1 and f2, where f1:f2 = m:n (Sauseng *et al.*, 2008; Siebenhühner *et al.*, 2016; Tass *et al.*, 1998). Cross-frequency phase synchrony may regulate neuronal communication at the speed of the higher frequency through consistent spike time relationships, whereas phase-amplitude coupling and amplitude-amplitude coupling operate on the time scale of the slower oscillation (Palva & Palva, 2012). Cross-frequency phase synchrony has been observed in human M/EEG data during rest (Jirsa & Muller, 2013; Nikulin & Brismar, 2006; Palva *et al.*, 2005; Sauseng *et al.*, 2008; Siebenhühner *et al.*, 2020) and attentional and working memory tasks (Akiyama *et al.*, 2017; Palva *et al.*, 2005; Sauseng *et al.*, 2008; Sauseng *et al.*, 2009b; Siebenhühner *et al.*, 2016), as well as in LFPs in the rat hippocampus (Belluscio *et al.*, 2012; Zheng & Zhang, 2013), and in human intracranial data during working memory task performance (Chaieb *et al.*, 2015) and at rest (Siebenhühner *et al.*, 2020). Critically, evidence points to phase-amplitude coupling and phase-phase-coupling/ cross-frequency phase synchrony being different CFC mechanisms as established by their differential contribution to working memory (Siebenhühner *et al.*, 2016) and distinct spectral and anatomical patterns at rest (Siebenhühner *et al.*, 2020).

Amplitude-amplitude coupling, in which the amplitudes of the fast and slow oscillations are coupled, has been observed in neuroimaging studies (Bruns & Eckhorn, 2004; de Lange *et al.*, 2008), but its functional relevance is less clear, since such coupling is independent of spike time relationships per se (J. M. Palva & S. Palva, 2018).

Other putative forms of CFC in which the frequency of the faster oscillation itself is modulated, like phase-frequency, amplitude-frequency, and frequency-frequency coupling have received less attention, partially due to methodological problems, e.g., challenges in determining instantaneous frequency peaks. Nevertheless, there have been a few studies that investigated such CFC forms using EEG (Jirsa & Muller, 2013) or intracranial recordings (Ray & Maunsell, 2010). More research here seems warranted, as frequencies – especially in the gamma band – are known to change during tasks, which may distort assessments of this type of CFC.

The exact relationship between different forms of CFC is still debated. Studies have shown evidence that phase-amplitude coupling and cross-frequency phase synchrony differ in their spectral and spatial patterns as well as functional relevance and likely fulfil distinct, complementary roles (Siebenhühner *et al.*, 2020; Siebenhühner *et al.*, 2016). Still, there is evidence that they may interact and influence each other (Hyafil *et al.*, 2015b) or that they may all be assessed with non-specific methods such as bispectrum and bicoherence (Jirsa & Muller, 2013). Similarly, the relationship between local CFC, inter-areal CFC and phase synchrony are not well understood.

A common critique is that observations of CFC might be spurious, caused by increases in SNR during tasks, or by non-sinusoidal or non-zero mean waveforms (Aru *et al.*, 2015; Cole & Voytek, 2017; Gerber *et al.*, 2016; Jones, 2016). Methods have been proposed to control spurious observation with graph-theory methods for inter-areal CFC (Lozano-Soldevilla *et al.*, 2016; Scheffer-Teixeira & Tort, 2016; Siebenhühner *et al.*, 2020), based on waveform/signal shape analysis (Cole & Voytek, 2019; Fabus *et al.*, 2022; Giehl *et al.*; Jensen *et al.*, 2016; van Driel *et al.*, 2015), time-frequency representation of the signal (Jurkiewicz *et al.*, 2021), or statistical signal models (Dupre la Tour *et al.*, 2017). Using these, one study concluded that there was no evidence for non-harmonic local phase-amplitude coupling in human MEG (Giehl *et al.*, 2021), while others reported that, in realistic modelling, harmonic content in nonsinusoidal oscillatory dynamics does not necessarily indicate spurious CFC (Dellavale *et al.*, 2020; Jensen *et al.*, 2016; Jurkiewicz *et al.*, 2021; Velarde *et al.*, 2019)

One common issue with mainly older studies of CFC is that those often-made a-priori assumptions about the involved frequencies, potentially overlooking important interactions of other frequency combinations. This can be overcome partially by either using wide ranges of frequencies and frequency ratios or by methods that inherently identify coupled frequencies in a data-driven manner (Sorrentino *et al.*, 2022; Volk *et al.*, 2018).

Finally, only few studies have attempted to study the causal direction of CFC. Particularly for phase-amplitude-coupling, the common implicit assumption is that the low frequency entrains the faster one. Yet, one study using Granger causality analysis reported that, in rat hippocampus, gamma amplitude drives theta phase (high-to-low directionality) (Nandi *et al.*, 2019) while another, using non-linear auto-regressive models (Dupre la Tour *et al.*, 2017), suggested low-to-high directionality in rodent hippocampus, but high-to-low in rodent striatum and human cortex, and other studies have reported evidence for phase-phase-coupling being bi-directional in human EEG (Munia & Aviyente; Popov *et al.*, 2018) and in macaque auditory cortex (Marton *et al.*, 2019). These studies indicate that both low-to-high and high-to-low directed CFC may occur in the brain and be related to top-down and bottom-up signalling, respectively.

## 1.4   Travelling waves

An additional dimension one needs to consider when investigating brain oscillations is their spatial organisation across the cortex, and in particular, the way they propagate. In contrast to standing waves that oscillate in place like a guitar string, travelling waves are like ripples on a pond: activity peaks that move across the cortex over time. An oscillatory travelling wave is defined as a smooth phase shift between recording locations (electrodes, sensors, contacts) in the direction of signal propagation, in a specific frequency band (Muller *et al.*, 2018). One can define a mesoscopic oscillatory travelling wave when it is constrained to a single brain area, while macroscopic travelling waves can span the whole cortex. The speed of propagation of mesoscopic waves is about 0.1-0.8 m/s, which is consistent with the speed of signals travelling through long-range unmyelinated horizontal fibres present in superficial cortical layers (II and III; for review, Muller *et al.*, 2018). For macroscopic travelling waves - those travelling across the cortical surface - not all publications report propagation speeds, but when mentioned these exceed the speed of mesoscopic waves, i.e., 1-10 m/s, potentially due to propagation along myelinated white fibres.

Oscillatory travelling waves have been observed as early as the 1930s (for review, Hughes, 1995), but their neurophysiological mechanism is still poorly understood. This is likely due to the technical difficulty to measure them in humans with the currently available techniques (Grabot *et al.*), especially at the mesoscopic scale (but see, Petras *et al.*, 2025). Additionally, traditional data analysis methods (e.g., trial averaging, coherence measures) are ineffective in capturing travelling waves because they assume that cortical activity is space-time dependent—an assumption contradicted by the widespread occurrence of travelling waves (Alexander *et al.*, 2015). As already proposed by Hughes (1995), travelling waves could support information transfer between cortical locations. Alternatively, they could organise the level of excitation/inhibition in neural populations in space and time to efficiently process information; otherwise known as the scanning hypothesis (Goldman *et al.*, 1949). The latter is consistent with previous attempts to understand how distant brain areas communicate, such as Communication Through Coherence. However, the term 'Oscillatory Travelling Wave' is perhaps more parsimonious since it effectively proposes a functional mechanism allowing several regions to synchronise their activity (Alexander *et al.*, 2019; Alexander & Dugué) and achieve long-range communication (Jacobs *et al.*, 2025).

Research on oscillatory travelling waves has regained interest in the last decade, particularly with respect to their potential role in cognition, including perception, attention (see section 1.1.a) and memory (see section 1.2.a and 1.2.b). This opens new lines of research, critical to the understanding of the link between brain functions and neural oscillations. Yet, future research will need to clarify their neurophysiological substrates and underlying mechanisms. One might even ask whether spatial propagation is an intrinsic feature of neural oscillations: are all brain rhythms travelling waves?

## 1.5   Resting-state rhythmic activity

The brain's activity at rest, i.e., when there is no task-related activity, exhibits highly structured spatiotemporal patterns (Deco & Jirsa, 2012), which reflect the functional architecture of cortical networks (Singer, 2013). These spatiotemporal patterns arise from local and synchronised activity of the network's constituent nodes and can be measured with BOLD fMRI and electrophysiological methods (i.e., EEG, MEG, ECoG). Investigating brain activity during rest provides valuable insights into the neural underpinnings of cognition, as there is consensus that task-relevant cortical dynamics are already reflected in ongoing, resting-state activity (Deco & Jirsa, 2012; Raichle, 2015).

The most prominent resting-state activity is likely the parieto-occipital alpha rhythm (Salmelin & Hari, 1994), which is often visible in raw electrophysiological recordings (Adrian & Matthews, 1934). Other large-scale, characteristic rhythmic activity across the cortex has also been described, such as frontal theta and temporal gamma (Frauscher *et al.*, 2018; Groppe *et al.*, 2013; Mellem *et al.*, 2017). Furthermore, local spectral 'fingerprints' have been revealed for distinct brain areas (Keitel & Gross, 2016; Lubinus *et al.*, 2021; Mahjoory *et al.*, 2020; Myrov *et al.*, 2024). These region-specific spectral profiles are characteristic combinations of endogenous rhythms, usually with more than one prominent rhythm, that appear remarkably consistent in groups of healthy participants (Keitel & Gross, 2016; Lubinus *et al.*, 2021). The concept of spectral fingerprints of brain areas is closely related to that of neural population-specific resonance frequencies (Keitel & Gross, 2016; Rosanova *et al.*, 2009). This suggests that at least some of the resting-state rhythms originate from individual oscillators. However, given that a brain area can exhibit multiple distinct rhythms, it is presumed that they can engage in different functional states over time, each of which may have an associated peak frequency (Keitel & Gross, 2016). This implies that what we measure as local resting-state rhythmic activity could also reflect connection phenomena (see section 1.1.b). It is currently unclear which rhythmic activity arises from individual (local) oscillators, and which reflects synchronised (network) activity between areas. In addition to a local spectral organisation, it has been suggested that spectral properties change gradually across the cortex, creating a spectral gradient (Mahjoory *et al.*, 2020; Zhang *et al.*, 2018). Here, spectral gradients were observed with region-specific peak frequencies (i.e., most dominant spectral activity in a brain area) decreasing from posterior to anterior parts of the brain. The notion of cortical gradients is well-established for structural features (e.g., neuron density, myelin content, cortical thickness) and putatively reflects a global cortical and hierarchical organisation.

In addition to the described local intrinsic brain rhythms, a different angle is to look at frequency-specific connections within and across brain areas, i.e., functional connectivity. These resting-state *networks* have originally been identified using BOLD fMRI (e.g., Damoiseaux *et al.*, 2006). More recently, emerging evidence has established the presence of electrophysiological frequency-specific oscillatory networks (Brookes *et al.*, 2011; Fusca *et al.*, 2023; Vidaurre *et al.*, 2018). These studies use different methods to find inter-area associations, such as amplitude envelope correlations (Bijsterbosch *et al.*, 2017; Brookes *et al.*, 2011), phase coupling (Vidaurre *et al.*, 2018) or spectral coherence (Nolte *et al.*, 2004). A problem when looking at electrophysiological measures of connectivity are spurious observations due to field spread or spatial leakage (Palva *et al.*, 2018), but suitable corrections have been suggested to overcome this issue (for an overview, see Colclough *et al.*, 2016).

Importantly, despite being generalisable across individuals, both local rhythmic activity as well as oscillatory networks at rest also show large inter-individual variability (Fusca *et al.*, 2023; Simola *et al.*, 2022), which can predict individual variation in cognitive performance (Baltus & Herrmann, 2016; Barnes *et al.*, 2016; Lubinus *et al.*, 2025). Furthermore, resting-state rhythmic activity can change in the context of functional and structural brain reorganisation, as demonstrated by differences in spectral fingerprints between normally sighted and congenitally blind individuals (Lubinus *et al.*, 2021), or by spectral changes associated with schizophrenia (Hua *et al.*, 2020), depression (Fernandez-Palleiro *et al.*, 2020), Alzheimer's dementia (Pusil *et al.*, 2019), and others.

Importantly, both local oscillations and large-scale oscillatory networks are influenced by many biological factors such as individual genetics (Leppaaho *et al.*, 2019; Salmela *et al.*, 2016; Simola *et al.*, 2022; van Pelt *et al.*, 2012), brain microarchitecture (Myrov *et al.*, 2024) and structural connectivity pathways (D'Andrea *et al.*, 2019). Recent research, in fact, shows that at the level of individuals, the spectral fingerprints are consistent within an individual, allowing identification of individuals with excellent accuracy (da Silva Castanheira *et al.*, 2021; Haakana *et al.*, 2024). The possibility of using functional resting-state data to identify individuals has led to some ethical concerns over participant anonymity and whether it is acceptable to deposit 'anonymised' resting-state datasets on openly accessible servers. To address this issue, the field would benefit from guidelines to mitigate misuse of data.

## 1.6    Interactions with other bodily rhythms

Research investigating the link between neural oscillations and human cognition has long cast a spotlight on the tight link between brain neurophysiology and behaviour, ultimately neglecting modulatory influences from body physiology. However, accumulating evidence shows that respiration (Kluger *et al.*, 2021), cardiac activity (Candia-Rivera *et al.*, 2023), pupil dynamics (Pfeffer *et al.*, 2022), and gastrointestinal signals (Azzalini *et al.*, 2019; Banellis *et al.*, 2024) influence neural excitability, arousal, and general cognition. These observations are motivating a paradigmatic shift in the study of neurocognitive functioning and encourage new research on the link with body-brain dynamics (Criscuolo *et al.*, 2022; Kluger *et al.*, 2024), from health to pathology.

### 1.6.a    Respiration

Human respiration is a continuous, rhythmic sequence of active inspiration and passive expiration (Fleming *et al.*, 2011). Key respiratory structures like the preBötzinger complex and olfactory bulb are intricately connected to both deep and superficial cortices (Yang & Feldman, 2018), thus forming bidirectional pathways between respiratory control and cognitive function. As neural oscillations in the cortex are conceptualised to reflect brain states and encode task-relevant information (Thut *et al.*, 2012), attention to respiration-brain coupling has increased. Extensive rodent work (Ito *et al.*, 2014; Tort *et al.*, 2018) has demonstrated that oscillations across the animal cortex are influenced by respiration, suggesting that functional roles of brain activity should consider interactions of neural and peripheral activity.

Both invasive (Zelano *et al.*, 2016) and non-invasive human studies (Kluger & Gross, 2021) have shown that nasal respiration distinctly modulates neural oscillations across a wide cortico-subcortical network. Mechanistically, cross-frequency coupling (Canolty & Knight, 2010) provides an intuitive implementation of this link between respiration phase and oscillatory power: During nasal breathing, the airstream triggers mechanoreceptors connected to the olfactory bulb, thereby initiating infraslow neural oscillations closely coupled to the respiratory rhythm. The phase of these slow oscillations then drives the amplitude of faster oscillations and propagates to upstream areas both within and beyond the olfactory system. This way, oscillatory power throughout the brain and across frequency bands may be coupled to the breathing rhythm.

Modulatory behavioural effects of respiration have been shown in perceptual (Johannknecht & Kayser, 2022; Kluger *et al.*, 2021), motor (Kluger & Gross, 2020; Rassler & Raabe, 2003), and cognitive tasks (Arshamian *et al.*, 2018; Perl *et al.*, 2019). Taken together, these studies provide strong evidence for breathing-related changes in neural signalling - e.g. critical brain states like excitability or arousal - which in turn translate into behavioural changes. Critical open questions remain as to whether the coupling of respiration, brain, and behaviour is functional rather than epiphenomenological, and the extent to which complex, higher-order interactions of bodily signals modulate the observed effects.

### 1.6.b    Pupil-linked arousal

Nuclei in the brainstem and basal forebrain form an arising activation system that regulates cortical arousal through neurotransmitters (Harris & Thiele, 2011; Hasselmo, 1995; Lee & Dan, 2012; Steriade, 1996). Most prominent projections to the cortex are the release of norepinephrine by the locus coeruleus, and of acetylcholine by the basal nucleus of Meynert. Tonic arousal levels change with behavioural state, e.g., remaining quiet versus moving (Crochet & Petersen, 2006; Niell & Stryker, 2010; Polack *et al.*, 2013). Optimal levels of phasic fluctuations in arousal that fall between detrimental hypo- and hyperarousal can influence behavioural performance (Aston-Jones & Cohen, 2005; McGinley *et al.*, 2015; Yerkes & Dodson, 1908).

Pupillometry, the non-invasive measurement of pupil size, is a viable way of studying the effects of neuromodulatory arousal on human cognition (Bradshaw, 1967; Hess & Polt, 1964; Kahneman & Beatty, 1966). Recent studies, combining pupillometry and MEG/EEG, provide  evidence that neuromodulatory arousal has distinct effects on different cortical regions, and rhythmic activity in different frequency bands in the human brain (Pfeffer *et al.*, 2022; Podvalny *et al.*, 2021; Radetz & Siegel, 2022) with effects on cognitive function (Dahl *et al.*, 2020; Podvalny *et al.*, 2021; Waschke *et al.*, 2019).

To date, there is no unified description of how and to what extent rhythms in various cortices couple to (pupil-linked) arousal, also owing to the fact that arousal dynamics may only have limited inherent rhythmicity. Arousal-based pupil dynamics typically occur in a range of several hundred milliseconds, tapering off towards a 'speed limit' of about 2-3 Hz (e.g., (McGinley *et al.*, 2015). Interestingly, the pupil also engages in its own resting rhythm, the Hippus, with a frequency of around 0.2 Hz (Bouma & Baghuis, 1971). Moreover, pupil dynamics in the Hippus frequency range increase in power when entering a resting state and adopting an inward attentional focus in contrast to monitoring the external sensory environment (Kluger *et al.*, 2024; Pomé *et al.*, 2020). Investigating how and when the Hippus emerges and couples with cortical rhythms and other cyclical physiological processes, e.g, respiration (Kluger *et al.*, 2024; Melnychuk *et al.*, 2021), may provide further insights into the role of (pupil-linked) arousal on cognition.

### 1.6.c Cardiac rhythms

The cardiac cycle is characterised by a cyclic alternation of ventricular contraction and relaxation, also named systole and diastole, which determines the heartbeat. Rather than being a regular metronome, the heart is a dynamic pacemaker (Shaffer *et al.*, 2014), influenced by the sympathetic and parasympathetic (vagus) nerves, and influencing neural dynamics (Candia-Rivera *et al.*, 2024). Heart-brain interactions are modulated by sympathovagal activity and have direct influence on the insula, amygdala, hippocampus, and cingulate cortices (Catrambone *et al.*, 2024; Kim *et al.*, 2019). A recent study by Jammal Salameh *et al.* (2024) extends the proposed heart-brain network by suggesting that the heartbeat can entrain neural population activity in olfactory bulb neurons via mechanosensitive ion channels. By coordinating local neural spike timing, the cardiac cycle may orchestrate neural dynamics across the cortex, from prefrontal cortex to the hippocampus (Jammal Salameh *et al.*, 2024).

Similarly to brain activity (Zoefel & VanRullen, 2017), cardiac fluctuations are thought to instantiate alternating time-windows of high- and low-excitability, ultimately impacting (self-)consciousness, perception, and cognition (Park & Blanke, 2019; Tallon-Baudry *et al.*, 2018). Thus, visual (Galvez-Pol *et al.*, 2020; Kunzendorf *et al.*, 2019) and somatosensory processing (Al *et al.*, 2021; Edwards *et al.*, 2009), visual attention (Pramme *et al.*, 2014), and interoceptive awareness (Herman & Tsakiris, 2021), are better during systole than diastole. Similarly, sensorimotor processing, cortical and corticospinal excitability are maximal in the systolic phase (Al *et al.*, 2021; Galvez-Pol *et al.*, 2020), and action initiation preferentially clusters in the systole (Galvez-Pol *et al.*, 2020). However, there is a lack of standardised protocols, opposing results and conceptual interpretations (Candia-Rivera *et al.*, 2024; Engelen *et al.*, 2023), currently limiting our understanding of the complex interface between heart-brain-behaviour.

### 1.6.d   Gastric rhythms

Muscle contractions of the stomach are controlled by rhythmic electrical activity at about 0.05 Hz generated by the interstitial cells of Cajal (Rebollo & Tallon-Baudry, 2022). This gastric rhythm can be noninvasively measured with electrodes as the electrogastrogram, which displays an increased amplitude during digestion. Recent neuroimaging studies reported that this gastric rhythm modulates brain hemodynamics (Choe *et al.*, 2021; Rebollo *et al.*, 2018). This effect occurs predominantly in sensory and motor areas but largely spares higher-order cognitive or transmodal brain networks (Rebollo & Tallon-Baudry, 2022). Gastric modulation of neural activity is thought to arise from interoceptive signalling along the vagus nerve and the spinal cord (Müller *et al.*, 2022). Primary entry points of these signals are subcortical nuclei such as the nucleus tractus solitarius from which they are distributed across the cortex via the thalamus (Mayer, 2011).

A recent MEG study showed that the amplitude of alpha rhythms in the parieto-occipital cortex and the right anterior insula is modulated by the phase of the gastric rhythms (Richter *et al.*, 2017). This effect accounts for about 8% of the variance in alpha amplitude. Directed connectivity analysis suggested that the gastric rhythm drives the modulation in neural alpha oscillations. While the consequences of this modulation for perception and cognition remain unclear, gastric rhythms should be considered when studying brain rhythms.

### 1.6.e   Circadian rhythms

Circadian rhythms shape all organisms, actively influencing nearly every aspect of physiology and behaviour to adapt to the 24-hour day-night cycle (in Latin, 'circa' means about, whereas 'dien' means a day). Circadian rhythms include processes ranging from physiological and homeostatic, to protein synthesis and DNA replication, as well as behavioural routines like the sleep-wake and feeding cycles. This periodicity represents an internal clock, regulated by slow environmental changes, such as variations in light intensity and temperature throughout a day. Such a bodily time-keeping mechanism is an evolutionary achievement that allows the organism to adapt to recurrent environmental changes, by optimising physiological and behavioural processes to external contingencies (Yerushalmi & Green, 2009). Thus, similar to other bodily rhythms (including brain oscillations), circadian rhythms are endogenous oscillators that interact with environmental and/or bodily rhythms by adjusting their phase, but are (in principle) independent of them.

The neural core of the circadian timekeeping system in mammals is in the suprachiasmatic nuclei (SCN; (Dibner *et al.*, 2010). Next to this central clock, nearly every cell in the body possesses some time-keeping mechanism and transmits information to SCN neurons to synchronise circadian physiology to geophysical time. The circadian cycle modulates neural activity, humour, and behaviour (Dibner *et al.*, 2010) via the HPA axis, thus regulating glucocorticoid hormones (Spiga *et al.*, 2014). Importantly, when internal clocks are altered, i.e., when the alignment between endogenous and exogenous cycles is modified, for example when changing time zones, fundamental periodicities are impacted: sleep-wake cycles, metabolism, hormone secretion, food intake, cortisol levels, energy, mood, and immune system efficiency (see for reviews (Finger & Kramer, 2021; Patke *et al.*, 2020). Since circadian rhythms have been identified in all organs, heart, stomach, and liver, an important question emerges: how does the brain with its own rhythmicities interact with the circadian timekeeping system?

The link of circadian functions to cognition and behaviour is an alluring, if ill-understood, one. Causal relationships are notoriously hard to establish, not least because cognitive and mental health disturbances acerbate chronobiological disturbances and vice versa. It will be an important step forward to understand the coupling of oscillations at vastly different time scales: the comparably slow, circadian rhythmicities in bodily tissue and in basic psychological phenomena, such as perception (e.g., (Obleser *et al.*, 2021), and the comparably fast, sub-second cycle length of neural oscillatory networks that are central to this manuscript. The next section provides some pointers to this by looking at ideas how physiological and brain rhythms are linked more generally.

### 1.6.f   Limitations and open questions on body-brain interactions

Evidence suggests that a holistic and systematic assessment of body-brain coupling can deepen our understanding into how we evaluate, perceive, and act in a dynamically changing environment (Criscuolo *et al.*, 2022; Kluger *et al.*, 2024). While this field of research is growing, several questions emerge: first of all, how to best quantify body-brain coupling? Are brain-based functional connectivity measures suitable for assessing body-brain coupling? Recent propositions postulate that a dynamical system approach offers the best way forward (Criscuolo *et al.*, 2022; Kluger *et al.*, 2024): body-brain interactions can be characterised by low-dimensional spatiotemporal states. This perspective embraces inter-individual variability in bodily and brain rhythms, delineates a tight link to variability in behavioural rhythms (Klimesch, 2018), and further promises to uncover valuable biomarkers for pathology (Banellis *et al.*, 2024; Kluger *et al.*, 2024). However, in the absence of empirical evidence, it remains unclear how to best dissociate 'optimal' from 'altered' body-brain dynamics, and their influence on neurocognitive functioning. A further question is whether it would be possible to modulate body-brain dynamics to influence arousal states and cognition (Criscuolo *et al.*, 2022).

## 2   Oscillatory mechanisms and their role in cognition

Many of the mechanisms and observed phenomena described above have been closely linked with cognitive functions. We currently assume that they provide the neural implementation of classical psychological concepts, such as attention and memory. Below we give an overview of our current understanding of these links, including highlighting open questions and debates.

### 1.7   Perception & attention

#### 1.7.a   Visual perception & attention

*Visual perception*

Brain rhythms, especially alpha, seem to index the momentary excitability of our visual cortex to incoming stimulation. Surprisingly, it is still not entirely clear how to interpret the effects of alpha oscillations that precede a stimulus on its detection: Does a state of stronger excitability, reflected by weak pre-stimulus alpha power, help observers to see targets better, or does it bias observers to report target presence?

Strong parieto-occipital alpha oscillations impede detection of visual targets (Ergenoglu *et al.*, 2004; van Dijk *et al.*, 2008) or perception of TMS-induced phosphenes (Romei *et al.*, 2008b; Samaha *et al.*, 2017a). More recent analyses based on signal detection theory, a theory of perceptual decision making, have shown that the effect of pre-stimulus power on detection is better characterised as an effect on bias, such that pre-stimulus states of heightened excitability (indicated by weak alpha power) induce a liberal bias to report stimulus presence (Iemi & Busch, 2018; Iemi *et al.*, 2017; Limbach & Corballis, 2016).

Nevertheless, pre-stimulus alpha power affects subjective stimulus visibility (Benwell *et al.*, 2022; Benwell *et al.*, 2017), subjective contrast appearance (Balestrieri & Busch, 2022), or confidence (Samaha *et al.*, 2017b), more in line with a perceptual bias, whereby states of strong neuronal excitability amplify the intensity and target-likeness of both sensory signals and sensory noise. Importantly, this interpretation links pre-stimulus alpha power to conscious access and read-out of information and suggests a role at later cortical and higher-order information processing stages (discussed in more detail in the section on visual attention below).

In contrast to alpha power, evidence for effects of alpha phase on perception is less equivocal (Keitel *et al.*, 2022). This is surprising, given that alpha phase is the original candidate for a shutter mechanism that implements the notion of 'windows of opportunity' or 'perceptual cycling' neurally. Although some findings support the perceptual cycling hypothesis (Busch & VanRullen, 2010; Kizuk & Mathewson, 2017; Spaak *et al.*, 2014; VanRullen, 2016), others do not (Benwell *et al.*, 2022; Benwell *et al.*, 2017; Melcon *et al.*, 2024; Ruzzoli *et al.*, 2019; van Diepen *et al.*, 2015). Methodological challenges in estimating phase in human EEG/MEG recordings can overshadow potential effects (Vigué-Guix *et al.*, 2022). Recognising intra- and inter-individual variability may critically shape such effects, offering a potential explanation for prior inconsistencies (Romei & Tarasi, 2025).

The 'pulsed inhibition' account (Mathewson *et al.*, 2009) encapsulates the idea that low-frequency oscillations, here alpha, provide the temporal scaffold for visual sampling. Given this assumed role of the phase and frequency of low-frequency oscillations in temporal scaffolding, high-frequency oscillations may play their part in what is being scaffolded. The idea of 'duty cycles', for example, posits that oscillatory activity in the gamma frequency range (30 - 80 Hz), nested into the alpha cycle, codes for different objects in the visual field (Jensen *et al.*, 2012), constituting a case of cross-frequency coupling (CFC) for visual perception (see section 1.3). Going into the more excitable period of the alpha cycle releases neuronal firing from inhibition and starts a sequence of activations of neuronal representations that manifests as gamma activity (Montemurro *et al.*, 2008).

The duty-cycle notion, especially the nested gamma, interfaces with a long-standing idea of how the visual system solves the hard problem of integrating different features (colour, orientation, shape) into coherent percepts of separate objects: Early theoretical suggestions by Milner (1974) and von der Malsburg and Willshaw (1981) ultimately found experimental support in recordings from the visual cortex of cats (Eckhorn *et al.*, 1988; Gray & Singer, 1989): Grating stimuli elicited coherent oscillatory activity in the gamma range between neuronal populations that preferred respective stimulus features. This phase-locking of high-frequency activity became widely regarded as the neuronal implementation of feature binding and, following further replications in animals (Engel *et al.*, 1991; Kreiter & Singer, 1996), was formalised in the 'binding-by-synchrony' hypothesis. Whether binding-by-synchrony sufficiently solves the feature binding problem in human vision remains an active area of research (Isbister *et al.*, 2018; Palanca & DeAngelis, 2005). A more recent view is that of high-frequency activity as a mechanism for predictive processing (see section 1.2.c), where highly predictable visual input appears to be associated specifically with high-frequency synchronisation in the visual cortex (Uran *et al.*, 2022).

The role of brain oscillations as a discrete sensory sampling mechanism can also be tested through peak frequency modulations, as initially proposed by Varela *et al.* (1981). Frequencies indicate the rate of change between inhibition and excitation, possibly representing a more stable measure of sampling than phase over time. In line with this, higher alpha peak frequencies translate into higher temporal accuracy when sampling information over several cycles, in both visual (e.g., (Samaha & Postle, 2015; Wutz *et al.*, 2018) and multisensory perceptual experiences (e.g., (Cooke *et al.*, 2019; Migliorati *et al.*, 2020) in some, but not all studies  (Gulbinaite *et al.*, 2019); see (Samaha & Romei, 2024), for a review). Manipulating peak alpha frequency via TMS, prior to the presentation of a peri-threshold stimulus, led to higher accuracy for faster peak frequencies (Coldea *et al.*, 2022; Di Gregorio *et al.*, 2022). Taken together, these results suggest that (alpha) rhythms influence perceptual experience by providing a better temporal resolution, or a more effective sampling per cycle for higher frequencies (also see (Tarasi & Romei, 2024).

*Visual attention*

We understand selective visual attention as a set of interacting mechanisms that aim at prioritising behaviourally relevant sensory input. Most of these mechanisms have been linked to rhythmic brain activity: facilitating relevant input, actively suppressing or filtering out irrelevant input, and the dynamics of (re)allocating the focus of visual attention. Requiring coordination across distributed brain areas, these mechanisms may also rely on large-scale synchronisation.

Allocating attention involves the frontal eye fields (FEF), a section of the premotor cortex, which highlights the close link between visual attention and overt gaze behaviour. When exploring a visual scene naturally, our gaze will saccade between scene elements with an average period of 0.2 sec, or a rate of 5 Hz (Otero-Millan *et al.*, 2008). This saccading provides 'fresh' sensory information at times of high cortical excitability determined by ongoing cortical low-frequency rhythms in the delta and theta ranges, according to the Active Sensing account (Schroeder *et al.*, 2010) and commensurate with the notion of rhythmic sampling (VanRullen, 2016). Understanding the intricate links between brain rhythms and oculomotor behaviour have recently become a research focus (Cruz *et al.*, 2025; Popov *et al.*, 2023).

Recent findings also suggest that covertly allocating attention relies on similar mechanics as overt gaze. For example, the accuracy of detecting peripheral targets varies with a period that falls into the theta frequency range (Landau & Fries, 2012). These variations were also found to be in anti-phase with an opposite peripheral location (Fiebelkorn *et al.*, 2013; Landau *et al.*, 2015) lending support to a gaze-like spatio-temporal sampling for involuntary shifts of attention and thought to be implemented through cortical low-frequency oscillations, phase-reset by salient cues that capture attention involuntarily.

Brain rhythms further subserve the voluntary allocation of attention. Cued shifts of visuospatial attention are associated with the modulation of alpha activity. This is typically shown in the form of alpha lateralisation, where higher alpha amplitude is observed in visual cortical regions that represent an ignored visual hemifield (e.g., (Thut *et al.*, 2006; Worden *et al.*, 2000). Originally interpreted as attention-related gain modulations in early visual cortical areas, these modulations are now predominantly understood as a gating mechanism, relaying only relevant visual input to higher-tier visual cortices (Jensen, 2024; Peylo *et al.*, 2021) such that could for instance be implemented via phase-synchronization among the attention-control network (Lobier *et al.*, 2018). In line with this, alpha lateralisation and attentional gain are largely unrelated in the early visual cortex (Antonov *et al.*, 2020; Gundlach *et al.*, 2020; Keitel *et al.*, 2019; Zhigalov & Jensen, 2020). More precisely, while the spontaneous alpha rhythm's inhibitory effect seems to apply in primary visual cortex (Iemi *et al.*, 2019), this inhibition does not appear to be under attentional control or at least does not translate into attentional gain modulations.

Oscillatory frameworks of visual attention remain largely focused on alpha activity. While this neglects the roles of other rhythms, a range of recent findings justify zooming in on alpha further: For example, parietal, occipital and temporal cortices show at least two distinct alpha rhythms (Barzegaran *et al.*, 2017; Keitel & Gross, 2016). Sokoliuk *et al.* (2019) also report two occipital alpha generators that show distinct functional characteristics: one linked to spatial attention allocation, the other to attentional effort or reflexive mechanisms (see also (Cruz *et al.*, 2025). Alpha rhythms have further been separated by their tendencies to show distinct (or combined) amplitude increases or frequency decreases over time (Benwell *et al.*, 2019), effects whose relative contribution to visual attention remain largely unexplored (Kopčanová *et al.*, 2025).

More recently, casting alpha as travelling waves (see section 1.4) has allowed further insights. Although travelling waves are a known concept (Klimesch *et al.*, 2007; Patten *et al.*, 2012), they have recently been re-discovered in the context of visual attention (Alamia & VanRullen, 2019; Fakche *et al.*, 2024; Lozano-Soldevilla & VanRullen, 2019). Alamia *et al.* (2023) showed that sets of alpha waves travel along the cortex on an anterior-posterior axis, forward and in reverse. Importantly, backward propagating waves increased in the hemisphere ipsilateral to a cued location, potentially implementing a gating of unattended visual input. Forward propagating waves, only found during visual stimulation, seemed to indicate a feedforward process instead. However, methodological challenges in analysing travelling waves remain to be addressed (Das *et al.*, 2022; Zhigalov & Jensen, 2023).

Voluntary shifts of attention seem under the control of fronto-parietal attention networks that have been shown to synchronise at alpha frequencies (Capotosto *et al.*, 2009; Sauseng *et al.*, 2011; Sauseng *et al.*, 2005). Extensive long-range connections exert controlling influences on occipital visual areas (Buffalo *et al.*, 2010; Debes & Dragoi, 2023). Involving a wider range of frequencies, these connections may selectively synchronise distant populations of neurons that represent attended locations or features by using beta- (Gross *et al.*, 2004; Hipp *et al.*, 2011) or gamma-range oscillations (Gregoriou *et al.*, 2009). More specifically, synchrony between frontal and parietal cortex is strong specifically in low frequencies during goal-driven attention, while it is strong in high frequencies during involuntary, stimulus-driven shifts (Buschman & Miller, 2007).

This also aligns with the idea that lower-frequency (alpha/beta) activity tends to carry predictions in the feedback direction, while higher-frequency (gamma) activity is associated with prediction errors carried forward (Bastos *et al.*, 2012). The winner-take-all mechanism for an attended stimulus, posited by the influential biased competition account of selective attention (Desimone & Duncan, 1995) can also be modelled with selective synchronisation, as described by 'communication through coherence' (CTC, (Fries, 2015)). However, it has recently been suggested that the gamma-based selective routing of information between brain areas can be captured in a framework in which coherence is a consequence rather than a requisite, calling into question the role of phase synchronisation in exerting attentional control (Dowdall *et al.*, 2023).

### 1.7.b   Auditory perception and attention

*Auditory perception*
The auditory cortical system exhibits distinctive rhythmic activity (spectral 'fingerprints') during rest (Section 1.5) and active listening (Keitel & Gross, 2016). Accordingly, the perception of acoustic events is also modulated by brain oscillations and their various properties (for review, see (Gourévitch *et al.*, 2020). Nevertheless, oscillatory mechanisms in auditory perception do not seem to be a mere copy of those found in its 'big brother' vision (for a comparative review, see (VanRullen *et al.*, 2014).

First, there is evidence that spontaneous brain oscillations do not phasically modulate the perception of an attended auditory target, as long as the timing of this target cannot be predicted (Lui *et al.*, 2025; VanRullen *et al.*, 2014; Zoefel & Heil, 2013). This contrasts with findings from the visual domain, although these are also debated (see Section on visual perception). A 10-Hz reverberation, or "perceptual echo" of sensory input, prominently observed for visual input and taken as a form of sensory replay (VanRullen & Macdonald, 2012), is also absent in audition (Ilhan & VanRullen, 2012). This difference between modalities has been explained by audition's need to cope with rapidly fluctuating input that can impede oscillatory 'sampling' in some scenarios (VanRullen *et al.*, 2014). Indeed, auditory phase effects can be reinstated by making acoustic targets irrelevant, in line with the notion that auditory oscillations are suppressed if critical sensory information can occur (and be 'lost') at the low-excitability phase (Lui *et al.*, 2025). Also, oscillatory phase effects on perception have been described for multi-modal stimuli that include an auditory component but might not originate from auditory cortex (e.g., (Leonardelli *et al.*, 2015; Thézé *et al.*, 2020). An intriguing additional possibility is that the phase of neural oscillations in the auditory system provides a temporal structure to stimulus representations: Neural populations, which process input that is more likely to occur are more sensitive and therefore active in earlier parts of a high-excitability phase, similar to the "duty-cycle" idea as originally proposed  for the visual and hippocampal systems (Jensen *et al.*, 2012; Lisman & Jensen, 2013). Ten Oever *et al.* (2024) found evidence for such a mechanism in the auditory system, where the phase of neural oscillations biased the perception of ambiguous speech depending on the likelihood of its constituents (phoneme and word frequencies (Ten Oever *et al.*, 2024).

Second, at least based on its impact in the corresponding fields, sensory entrainment (Section 1.1.c) is a more important process in the auditory than in the visual system. Entrainment has been proposed to play a fundamental (Giraud & Poeppel, 2012) and causal role (Riecke *et al.*, 2018; Zoefel *et al.*, 2018a) in auditory and speech processing. There is an ongoing debate on whether neural entrainment involves endogenous brain oscillations (Section 1.1.c; (Atanasova *et al.*, 2025; Duecker *et al.*, 2024; Haegens & Zion Golumbic, 2018; Keitel *et al.*, 2014; Zoefel *et al.*, 2018b), but some evidence for this involvement has been reported, such as through modelling of music (Doelling *et al.*, 2019), and entrainment to speech rhythm (Kösem *et al.*, 2018; van Bree *et al.*, 2021). Auditory entrained responses and their sustained effects are stronger for certain stimulus rates (Farahbod *et al.*, 2020; L'hermite & Zoefel, 2023; Teng & Poeppel, 2020), further supporting the notion of entrained endogenous oscillations (but see (Atanasova *et al.*, 2025). Recent research (L'hermite & Zoefel, 2023) has also suggested that a regular presentation of auditory events can lead to reduced perception in phase with the auditory stimulus (rather than improved perception as predicted by initial theories of entrainment; (Lakatos *et al.*, 2008), possibly reflecting habituation at specific time points and sound frequencies (Costa-Faidella *et al.*, 2011). Future research needs to test whether entrainment and habituation are competing mechanisms, and in which situations they occur, especially because in-phase perception can be improved in some experimental settings (for review, see (Haegens & Zion Golumbic, 2018).

An important concept in studying auditory perception by means of entrainment is that the stimulus regularity leads to predictable moments of stimulus presentation. This temporal predictability allows the alignment of high-excitability phases to the expected informative moment in the stimulus (Schroeder & Lakatos, 2009). The notion of 'active sensing' is conceptually related (see section 2.1a), and states that audition actively allocates neural resources to the expected timing of upcoming events (Schroeder *et al.*, 2010). Importantly, active sensing is not restricted to regular stimulation - the timing of non-rhythmic events can also be predictable - and therefore goes beyond entrainment. An important role of auditory-motor interactions has been proposed in such a context (Morillon & Schroeder, 2015), where the motor system's beta and delta oscillations 'prepare' audition for anticipated events through phase interactions with auditory areas Morillon (Morillon *et al.*, 2019; Morillon & Baillet, 2017).

Auditory oscillations can be produced or 'reset' by acoustic events like noise, and then fluctuate briefly in the delta-theta range Ho (Ho *et al.*, 2017; Kayser, 2019). Moreover, gamma oscillations seem linked to the temporal resolution of auditory processing (Baltus & Herrmann, 2016). These findings suggest that auditory 'sampling' at specific rates (~delta/theta and gamma) does exist, but it might need to be evoked through sensory input that entrains or resets these brain rhythms. Together, complex interactions between sensory and neural dynamics might explain why oscillations sometimes but not always play a role in the auditory system, and further studies are required to understand these interactions.

Third, alpha oscillations in the auditory EEG do not seem as prominent as in vision (Weisz *et al.*, 2011). This might be due to the anatomy or smaller size of auditory cortices that makes their activity more difficult to measure (Weisz *et al.*, 2011). However, putatively entrained auditory activity is prominent in the EEG (Obleser & Kayser, 2019), suggesting that anatomical reasons are not sufficient to explain the prominence of alpha oscillations in vision. Nevertheless, there is little doubt that alpha oscillations do exist in the auditory cortex and play a role for auditory processing (Billig *et al.*, 2019), including auditory attention (see next section; Wöstmann *et al.*, 2021). Recent work in both non-human primates (Lakatos *et al.*, 2019) and humans (Kasten *et al.*, 2024) suggests that alpha oscillations in audition reflect an absence of attention to sensory information and exhibit slow regular alterations with periods of reduced alpha oscillations and enhanced external attention. Taken together, the role of alpha in auditory perception is not as well understood as it is in vision (Section 2.1.a), and additional research should aim at a better characterisation.

*Auditory attention*

Relevant sounds are often masked, in time and frequency, by other distracting sounds. How does the brain select relevant target sounds? Research has shown that selective auditory attention is no unitary mechanism but composed of intertwining sub-processes (e.g., auditory object formation and object selection; (Shinn-Cunningham & Best, 2008)), and involves subcortical structures (Wimmer *et al.*, 2015), as well as regions across temporal (Mesgarani & Chang, 2012) and frontal cortices (Besle *et al.*, 2011). Previous work suggests at least three broad classes of auditory attentional phenomena which are potentially implemented by neural oscillations.

First, low-frequency (esp. delta/theta, approx. 1–8 Hz) neural oscillations have been proposed to phase-align to auditory signals, such that optimally excitable auditory-cortical states align to the rhythmic (i.e., often predictable) structure of the attended sounds ('entrainment in the narrow sense', (Obleser & Kayser, 2019). Specifically, in non-human primates it has been demonstrated that slow oscillations in the local field potential (LFP) in layers of primary auditory cortex become entrained to attended sound (Lakatos *et al.*, 2013). For complex stimuli such as speech (using M/EEG), a seemingly very similar phase-locking between the speech envelope and the neural response from auditory cortical sources ('neural tracking' or 'speech tracking', see section on auditory perception above) is also enhanced for attended versus ignored speech (Ding & Simon, 2012). Here, an important caveat is that speech tracking and its attentional modulations are not easily proven to be true neural oscillatory phenomena but likely involve a series of stereotypical, evoked responses to acoustically present 'edge' events or linguistically imposed segmental or phrasal boundaries (e.g., (Oganian *et al.*, 2023). This is an open issue awaiting to be resolved (also see section above, section 1.1b). However, it is possible that one of the most prominent electrophysiological signatures of auditory attention – i.e., modulation of phase-locked auditory evoked potentials by attention (Roth *et al.*, 1970) – might at least in part rest on a true phase-reset of slow neural oscillations (Makeig *et al.*, 2002; Sayers & Beagley, 1974).

Second, the power of alpha (~10 Hz) oscillations relates to cortical inhibition and has been implicated in the enhancement of targets (low alpha) and suppression of distraction (high alpha). It is long known that the power of neural oscillations in the alpha band is modulated by auditory attention (Adrian, 1944). Relatively suppressed alpha power in task-relevant cortical regions and enhanced power in task-irrelevant regions have been associated with target enhancement and distractor suppression in auditory attention, respectively (Schneider *et al.*, 2022; Strauss *et al.*, 2014). However, other research has implicated increased inter-areal alpha oscillations in auditory cortex with attentional functions (e.g., Bollimunta *et al.*, 2008). Alpha power modulations are prominent signatures of auditory attention, as they have been shown to reflect auditory spatial (Wöstmann *et al.*, 2016), temporal (Wöstmann *et al.*, 2020), and object-based attention (de Vries *et al.*, 2021). Although cortical surface maps in M/EEG often show a mixture of parietal, occipital, and temporal contributions to alpha power modulation in auditory attention, recent evidence from electrocorticography suggests that auditory cortical regions host sound-specific alpha oscillators (Billig *et al.*, 2019) and that alpha oscillations during an auditory task are suppressed in locations within the auditory system (de Pesters *et al.*, 2016).

Third, the power of auditory-induced gamma oscillations (> ~40 Hz) is enhanced for attended sound (Ray *et al.*, 2008). As opposed to alpha power, sound-induced gamma power is thought to reflect active auditory processing (Crone *et al.*, 2001). Salient distraction has been shown to suppress target-related gamma responses (Huang & Elhilali, 2020), potentially reflecting limited attentional resources. Supporting evidence comes from studies showing that higher gamma power was related to better auditory attention performance (Ahveninen *et al.*, 2013), and that stimulation of lateralised gamma oscillations (as compared to stimulation of lateralised alpha oscillations) modulated listeners' accuracy in an auditory spatial attention task (Wöstmann *et al.*, 2018).

Several challenges remain in understanding the putative neural oscillatory mechanisms in auditory attention. First, although research has unveiled a plethora of oscillatory phenomena related to auditory attention, it is often unclear how these (if at all) are relevant to attentional selection behaviourally. That is, only if an enhanced neural response to an attended stimulus explains better target selection (e.g., higher speech comprehension scores), can it be considered functionally relevant to auditory attention.

Second, we often lack specification in time and origin of neural oscillations involved in auditory attention. This can lead to seemingly contradicting results (e.g., alpha power increases in one study but decreases in another), because the underlying neural oscillators are topographically distinct (e.g., one study found modulation of parietal and the other of temporal alpha power). Note that such a lack of topographical specificity also challenges characterisation of auditory-control neural networks which were found to underlie behavioural performance in challenging tasks (e.g., (Alavash *et al.*, 2021).

Third, although some attempts have been made to dissociate modulations of neural oscillations associated with different sub-processes of auditory attention (e.g., target enhancement versus distractor suppression; (Wöstmann *et al.*, 2019)) the field must adapt their paradigms and analyses to be able to derive precise associations of neural oscillatory mechanisms with sub-processes of attentional selection (Wöstmann *et al.*, 2022).

Finally, while body movements constitute an important mechanism to control the sensory sampling of the environment (i.e., enhanced sampling of targets and avoidance of distraction), they are restrained or considered a confound in most auditory attention research (but see (Kondo *et al.*, 2012)). Only if we understand how neural oscillations interact with body movements can we understand their role for auditory attention.

### 1.7.c   Multisensory perception & attention

Processing sensory input provided by multiple channels requires a fine-tuned mix of weighing, integrating and segregating of auditory, visual, tactile and other input. For example, seeing a speaker's lips aids understanding speech in noisy environments (e.g., (Begau *et al.*, 2021; Sumby & Polack, 1954). Conversely, focusing on a visual task, such as reading, usually requires filtering out potentially distracting auditory information (Vasilev *et al.*, 2019). Considering oscillatory processes as the underlying neurophysiological implementation has advanced our understanding of how multisensory processing plays out in the brain (Keil & Senkowski, 2019).

A question that persists is how multisensory percepts, i.e., stimuli composed of visual, auditory, and features from other modalities, form and are represented neurophysiologically. Bizley *et al.* (2016) have argued for a layered hierarchy where 'integration' applies to any instance where input in one sensory modality impinges on perceptual processing in another. 'Binding', in contrast, would be a special form of integration that creates unified multisensory percepts. They suggest that interactions between early cortical areas provide the neural substrate for binding, which aligns with findings on how senses interact through oscillatory activity: cross-modal phase resets (Lakatos *et al.*, 2009; Mercier *et al.*, 2013) and phase synchrony (Senkowski *et al.*, 2008).

Providing an 'integration' substrate, Lakatos *et al.* (2009) showed that a salient stimulus will phase-reset low-frequency rhythmic activity not only in cortical areas processing the stimulus modality, but also in cortices processing different modalities (also see (Mercier *et al.*, 2013), via direct (monosynaptic) cortico-cortical connections (Falchier *et al.*, 2010). This transient alignment of perceptual cycles across senses will facilitate the processing of temporally and possibly spatially co-occurring stimuli (Lakatos *et al.*, 2009) and may explain why a visual search becomes trivial when changes in target appearance co-occur with a sound (Van der Burg *et al.*, 2008). Moreover, the common temporal scaffolding provided by crossmodal phase resets may afford cortico-cortical coherence in beta and gamma frequency ranges, which could be the neural substrate of 'bound' multisensory stimulus representations (Senkowski *et al.*, 2008). In the visual modality, the binding strength of two object features shows a natural periodic co-fluctuation at low frequencies (Nakayama & Motoyoshi, 2019). Finding a similar fluctuation in the power of high-frequency neural activity representing the constituent unisensory features of a multisensory object would support the vital role of low-frequency oscillatory phase in multisensory integration.

The relevance of crossmodal temporal scaffolding for cognition has also been probed with rhythmic sensory stimulation in one modality, then testing for corresponding periodicities in perceptual judgments in another. Several findings supported such effects of 'crossmodal entrainment' (Albouy *et al.*, 2022; Bauer *et al.*, 2021). In audiovisual speech, for example, sustained multisensory input streams share (quasi-)periodic temporal dynamics in lower frequency ranges that allow for crossmodal predictions about upcoming content (Biau *et al.*, 2021; Crosse *et al.*, 2015). However, other findings challenge the notion of crossmodal entrainment (Pomper *et al.*, 2023). Barne *et al.* (2022) add that its effects may be rather general in that predictions of the modality of upcoming stimulation aid in pre-activating respective sensory cortices.

The relative weighting of modality-specific input likely requires instances of supramodal control that may be implemented via top-down connections from higher-order to early sensory cortices (Talsma *et al.*, 2010) and much resemble the gating mechanism discussed for visual attention (section 1.2.b). In fact, low-frequency, and in particular alpha rhythms, may play a very similar role in multisensory processing, and experimental findings support a cross- or supramodal gating mechanism (Bauer *et al.*, 2012; Mazaheri *et al.*, 2014). Insights remain limited however as studies incorporating stronger stimulus competition, thus creating the demand for attentional filtering, in multisensory processing are scarce, and the role of attention is typically investigated by cueing one modality in a multisensory stimulus or task (but see e.g. (Begau *et al.*, 2022)).

An additional locus of control seems to be triggered when attention needs to be divided between senses: Studies have reported increased fronto-central or parietal theta-range oscillations in participants who were monitoring unrelated sound and visual sequences (Keller *et al.*, 2017; McCusker *et al.*, 2020). This has been argued to link to the role of theta oscillations in cognitive control and the increased demands to keep track of unrelated, potentially conflicting sensory streams, however, without detailing the underlying processes with which this theta modulation influences multisensory processing.

## 1.8  Memory

### 1.8.a   Working memory

Working memory (WM) is the ability to hold relevant information in an active, accessible state over a brief time after this information is no longer physically present — this function presents a core element of human cognition (cf., (Baddeley *et al.*, 2015). The potential role of neural oscillations across different frequency bands has been a key driving force in promoting our understanding of working memory.

The relevance of alpha oscillations for WM has been highlighted by the finding that alpha power generally increases during memory maintenance, and further increases with memory load (Busch & Herrmann, 2003; Jensen *et al.*, 2002). A widely accepted hypothesis is that alpha oscillations index the disengagement of sensory areas, thereby protecting internal representations from interference by distracting input (Cooper *et al.*, 2003; Jensen & Mazaheri, 2010; Klimesch *et al.*, 2007). In line with this interpretation, Bonnefond and Jensen (2012) reported  an increase in alpha power with the expectation of a distractor during the maintenance interval. It should be noted though that the directionality of alpha modulations is somewhat inconsistent across studies, sometimes showing an attenuation rather than an amplification (reviewed by (Pavlov & Kotchoubey, 2022). Reconciling these different results, it has been proposed that alpha increases in sensory areas (i.e., supporting disengagement and distractor inhibition) when abstract or non-sensory information is retained, whereas it decreases to support retention when precise sensory features are task-relevant (van Ede, 2018).

Another debate applies to lateralised alpha activity during the maintenance interval. When to-be-remembered and to-be-ignored items appear in opposite visual fields, alpha power typically increases in the hemisphere processing the distractors compared to the hemisphere processing the targets (e.g. (Sauseng *et al.*, 2009b). This lateralisation has been interpreted as reflecting both inhibitory (distractor-related) and facilitatory (target-related) mechanisms. However, the effect remains ambiguous: it is often unclear whether the lateralisation reflects a true increase in alpha power ipsilateral to the target, a decrease contralaterally, or both (Schneider *et al.*, 2019). A few recent studies were able to differentiate between a target-related decrease and a distractor-related increase in alpha power (Poch *et al.*, 2018; Vissers *et al.*, 2016), supporting the notion that the removal of information from working memory is mediated by an inhibitory mechanism (i.e., a distractor-related increase). However, this contrasts with findings showing that alpha lateralisation does not vary with the number of items that become irrelevant following a retro-cue (Hakim *et al.*, 2021), nor with the relevance of distractors presented during the delay period (Noonan *et al.*, 2018). Therefore, the exact functional role of alpha oscillations in working memory continues to be controversial. Resolving whether alpha power modulations are related to active versus automatic inhibition (Noonan *et al.*, 2018) as opposed to or in addition to mnemonic prioritisation (e.g., van Ede, 2018) and how that relates to the concepts of temporary versus permanent removal (Lewis-Peacock *et al.*, 2018) requires a neutral baseline against which power increases and decreases can be meaningfully assessed (Schneider *et al.*, 2022).

Working memory typically stores not only one but multiple items. This poses two computational and theoretically relevant challenges: First, the different features constituting an individual object must be bound, while keeping them separate from other objects. Second, not only the identity of single items but also the spatial relation or sequential order between the individual items must be retained. Oscillations have been proposed to play a critical role in solving both problems:

Multi-item working memory relies on the ability to link related features constituting a perceptual object. On a perceptual level, this binding problem largely refers to the issue of how individual features of an object are integrated into a coherent percept, whereas in working memory, it additionally relates to the question whether working memory capacity is limited by the number of individual features or the number of bound objects (for a review of recent behavioural evidence, see (Schneegans & Bays, 2019). In terms of its neural basis, feature binding implies that feature-specific processing in anatomically distributed areas needs to be coordinated in a way that allows for the formation of neuronal assemblies that represent perceptual objects. The influential binding-by-synchrony hypothesis postulates that this is achieved through synchronised oscillatory gamma activity (e.g., (Basar-Eroglu *et al.*, 1996), see also section 2.1.a). In line with this theory, (Honkanen *et al.*, 2015) observed a local increase in gamma power for the storage of feature-bindings compared to the storage of individual features in visual working memory. Further support for the importance of gamma power for binding in working memory comes from studies using non-invasive brain stimulation (Tseng *et al.*, 2016) or multisensory integration (Senkowski *et al.*, 2009). However, there is also evidence that theta oscillations are important for multisensory working memory processes (Seemüller *et al.*, 2012; Xie *et al.*, 2021), and that alpha oscillations are causally involved in feature binding (Zhang *et al.*, 2019). Therefore, extensive experimental testing is still required to (i) specify the role of different frequency bands for unimodal as well as cross-modal binding (e.g., (Arslan *et al.*, 2025) as well as (ii) to clarify how storage of feature bindings is sustained during the retention interval (Pagnotta *et al.*, 2024).

Once object features are linked, how do we keep them apart from disparate objects, while retaining spatial relations and sequence information? Computational models suggest that multiple items are stored by nesting fast rhythmic brain activity (in the gamma range) into slower (theta) activity (Lisman & Idiart, 1995; Van Vugt *et al.*, 2014) (see also section 1.3 on cross-frequency coupling). These models predominantly differ in how individual items are represented and how this relates to working memory capacity limits. The original model (Lisman & Idiart, 1995; Lisman & Jensen, 2013) assumes that each item is represented by a gamma cycle, while multiple gamma cycles are aligned to different phases of a slower theta rhythm. This limits working memory capacity to the number of gamma cycles that can be nested into one theta cycle. Supporting evidence comes from human intracranial hippocampal recordings (Axmacher *et al.*, 2010; Chaieb *et al.*, 2015), EEG recordings over posterior-parietal scalp sites (Sauseng *et al.*, 2009b) as well as non-invasive brain stimulation studies (Wolinski *et al.*, 2018). However, in addition to such a nested (phase-amplitude) relationship, multiplexing of memory representations could also be achieved by cross-frequency phase-phase synchronisation (Siebenhühner *et al.*, 2016).

Conversely, Van Vugt *et al.* (2014) postulated that single items are represented by gamma bursts. Therefore, with each theta cycle only one item is represented; and working memory capacity is limited by the fact that items need to be reactivated after a few theta cycles in order to not lose their representation. As discussed recently (Sauseng *et al.*, 2019), the alternative framework is still compatible with evidence from non-invasive brain stimulation research, such that a longer theta cycle would increase the duration of a gamma burst, thereby increasing memory fidelity. Yet, this relationship still awaits empirical testing and could potentially provide an avenue for linking theta-gamma coupling phenomena to models of working memory that assume a flexible distribution of resources rather than fixed slots. Recent computational work has further proposed an extension of the original Lisman-Idart model, in which theta-gamma interactions occur in spatial modules and oscillatory inputs are conceptualised as travelling waves (Soroka *et al.*, 2024) (see also section 1.4).

While the above-mentioned framework mainly links posterior (and hippocampal) theta oscillations to the representation of working memory contents, theta activity in prefrontal and anterior cingulate cortex, particularly, has been associated with control of working memory processes (e.g., (Berger *et al.*, 2019; Riddle *et al.*, 2020). To date, dissociating between those two functions of theta activity has been difficult. Recently, however, Ratcliffe *et al.* (2022) demonstrated that working memory content can be decoded from posterior theta activity, whereas anterior theta oscillations were associated with coordination of this retained information in working memory. In addition, control of visual working memory has been associated with travelling alpha waves with different kind of inhibitory control being reflected by different travelling directions (Zeng *et al.*, 2024). Future research needs to establish whether rhythmic brain activity is truly essential for short-term retention of information at all or if storage of multi-item working memory could also be achieved without the involvement of brain oscillations.

### 1.8.b   Long-term memory

As a highly plastic region, the hippocampus plays a fundamental role in long-term memory (Bliss & Lomo, 1973; Scoville & Milner, 1957). Theta and gamma oscillations in the hippocampus have been suggested to coordinate spike timing for synaptic weight changes, thus contributing to memory formation (Düzel *et al.*, 2010; Hanslmayr *et al.*, 2016; Herweg *et al.*, 2020; Jutras & Buffalo, 2010).

Empirical and computational evidence show that theta oscillations, as a dominant signal in the hippocampus, provide time windows for synaptic plasticity and organise the dynamics between memory encoding and retrieval by timing the two processes occurring at opposing theta phases (Hasselmo, 2005; Kerrén *et al.*, 2018; Ter Wal *et al.*, 2021). Rodent studies suggest that gamma oscillations separate encoding and retrieval as theta phases do (Colgin *et al.*, 2009). Supported by evidence in human single neuron and intracranial EEG studies, fast gamma (> 65 Hz) activity increases specifically during successful episodic memory encoding (Griffiths *et al.*, 2019). In contrast, slow gamma (~25 – 50 Hz) power increases during successful episodic or spatial memory retrieval (Griffiths *et al.*, 2019; Vivekananda *et al.*, 2021).

Gamma oscillations operate at a time scale that is consistent with the narrow time window within which one of the most prominent plasticity mechanisms operates (~25 ms), spike-timing-dependent plasticity (Bi & Poo, 1998; Fell & Axmacher, 2011; Jutras & Buffalo, 2010). Therefore, gamma synchronisation might reflect effective synaptic plasticity during memory encoding, which is supported by the findings that successful memory formation is linked with increases in gamma power (Gruber *et al.*, 2004; Hanslmayr *et al.*, 2009; Long *et al.*, 2014; Sederberg *et al.*, 2007). Since both gamma synchronisation and theta phases are linked to synaptic modifications, stronger locking of gamma oscillations to optimal theta phases should have add-on effects for memory formation. This is reflected in stronger theta-gamma phase-amplitude coupling (see Section 1.3) during the integrated encoding of multiple items into long-term memory (Köster *et al.*, 2018; Staudigl & Hanslmayr, 2013).

As mentioned in Section 2.2a, computational models on theta-gamma phase-amplitude coupling suggest that gamma oscillations may reflect the activity of smaller cell assemblies that represents a particular item in a sequence of items (i.e. the digit '7' in a string of 1-5-9-7-3). The activity of each assembly increases at a specific phase of the theta cycle, thus a sequence of information can be coded by different assemblies being active at different theta phases (Lisman & Jensen, 2013). It is important to note here that, whilst each item can be represented in a different cell assembly, it is the locking of each assembly's gamma activity to different theta phases that neatly organises the activity of these assemblies in time. Such a mechanism supports whether the order of a sequence of events can be correctly remembered (Heusser *et al.*, 2016). Since most real-world events unfold on a temporal scale that is slower than the gamma range, the theta-gamma phase-amplitude coupling would induce spike-timing-dependent plasticity to form neural sequences, called synfire chains, then store a sequence of items into long-term memory (Qasim *et al.*, 2021; Reifenstein *et al.*, 2021; Skaggs *et al.*, 1996).

Hippocampal theta- and gamma rhythms have been a focus of human long-term memory research because of well-established theoretical models and empirical evidence found in animal studies. Although similar evidence has been found in the human brain, there are some discrepancies. First, human endogenous theta in the hippocampus is much less rhythmic, has a broader band and is smaller in amplitude than in animals (Qasim *et al.*, 2021; Watrous *et al.*, 2013). The presence of a strong theta rhythm may not be as crucial in humans as it is in rodents, especially since phase coding can occur independently of a consistent frequency. In fact, learning—both spatial and non-spatial—has been shown to rely on phase coding even in the absence of a regular hippocampal theta rhythm (Bush & Burgess, 2020; Eliav *et al.*, 2018; Qasim *et al.*, 2021). A recent study used periodic theta-range stimulation (see Section 1.1c) to manipulate stationary theta phase synchronisation between sensory regions with the aim to improve episodic memory through sensory theta entrainment (Clouter *et al.*, 2017). Their positive results were challenged by non-replications highlighting the role of trial-by-trial or inter-individual variability in entraining endogenous hippocampal theta dynamics (Serin *et al.*, 2024; Simeonov & Das, 2025; L. Wang *et al.*, 2018).

Questions remain about the role of theta power in memory formation. Whereas some studies find that theta power increases support memory formation (Hanslmayr & Staudigl, 2014; Herweg *et al.*, 2020) others show that decreases in hippocampal theta power can also be linked to successful subsequent memory (Burke *et al.*, 2013; Crespo-García *et al.*, 2016; Long *et al.*, 2014; Sederberg *et al.*, 2007; Staudigl & Hanslmayr, 2013). Power decreases may reflect increases in capacity for information coding that aid the formation of rich memory representation in neocortex (Hanslmayr *et al.*, 2016; Hanslmayr *et al.*, 2012; Michelmann *et al.*, 2016). Reduced synchrony among neurons, i.e., less correlated activity, likely underlying these power decreases, may allow for a greater diversity of neural responses, thereby enhancing the brain's capacity to encode rich and detailed information. A remaining question is if hippocampal theta power decreases and neocortical alpha/beta desynchronisation reflect common processes. Future research should investigate if decreases in theta power are due to a different reference scheme (Herweg *et al.*, 2020) or a downstream effect caused by neocortical alpha/beta desynchronisation, potentially reflecting increased sensory intake and processing.

Mixed findings regarding memory-related theta power changes may also be linked to the existence of (at least) two different theta activities, slow theta at ~3 Hz, which is more analogue to rodents' hippocampal theta at ~8 Hz (Jacobs, 2014), and a faster human theta rhythm at ~8 Hz. Slow theta is mostly prevalent in the anterior hippocampus and increases in power during successful memory formation, while fast theta is more prevalent in the posterior hippocampus and decreases in power during successful encoding (Goyal *et al.*, 2020; Lin *et al.*, 2017). A functional distinction between posterior fast theta and anterior slow theta may also be linked to the different directions of information flow for memory encoding and retrieval (Linde-Domingo *et al.*, 2019): Theta (as well as alpha) travelling waves (see Section 1.4) propagate from posterior to anterior cortex during memory encoding, whereas during retrieval, they propagate in the reversed direction (Mohan *et al.*, 2024; Muller *et al.*, 2018; Zhang *et al.*, 2018).

In future research, invasive and non-invasive brain stimulation studies can help elucidate the causal roles of theta rhythms, theta-gamma coupling and alpha/beta desynchronisation in memory formation or retrieval (Clouter *et al.*, 2017; Ezzyat *et al.*, 2017; Hanslmayr & Staudigl, 2014; Hebscher & Voss, 2020; Lara *et al.*, 2018; D. Wang *et al.*, 2018). Knowing which memory processes to target, e.g. encoding or retrieval or cognitive processes, such as semantic encoding or non-semantic encoding, as well as computational modelling of the underlying neuronal processes, will also help in setting stimulation frequencies more precisely. Also, stimulation parameters may need to be adjusted dynamically to individual hippocampal theta activity for successfully studying memory encoding in theta entrainment paradigms (Wang *et al.*, 2024).

## 1.9   Communication

### 1.9.a   Speech and language processing

*Speech processing and entrainment*

Speech recognition poses a non-trivial challenge for the brain. From the continuous speech signal, the corresponding linguistic units such as phonemes, syllables, phrases or sentences need to be identified. Theoretical proposals suggest that brain rhythms in the human auditory cortex (and other areas) entrain to quasi-rhythmic occurrences of linguistic units (see section 1.1c), allowing for speech segmentation. Particularly, the most prominent slow quasi-periodic acoustic energy fluctuations at the syllabic scale  (~4-7 Hz, theta band) are considered as cues for the alignment of theta brain rhythms during syllabic segmentation (Ghitza, 2011; Giraud & Poeppel, 2012; Gross *et al.*, 2013). Slow energy fluctuations can be measured in the acoustic envelope, which contains phonemic and syllabic transitions related to energy changes. Such energy changes, including rapid increases in acoustic energy or the energy peaks related to mid-vowels (which provide stable cues in noisy environments), may provide acoustic landmarks at the cochlear level for neural entrainment of brain rhythms through phase-resetting (Aubanel *et al.*, 2016; Doelling *et al.*, 2014; Ghitza, 2013; Gross *et al.*, 2013; Oganian & Chang, 2019). Although speech has no periodic but rather a quasi-rhythmic structure, computational models provide evidence that oscillator models are capable of tracking such quasi-rhythmicity (Doelling *et al.*, 2023; Pittman-Polletta *et al.*, 2021; Ten Oever & Martin, 2021).

At the same time, prominent acoustic edges elicit evoked responses that overlays oscillatory activity. Accordingly, there is an ongoing debate whether oscillatory entrainment is involved in syllable segmentation or whether the tracking merely reflects a succession of evoked responses (see also section 2.1.b). Empirical evidence suggests that speech tracking can be explained solely by an evoked-responses account (Oganian *et al.*, 2023). However, in line with an entrainment interpretation, several studies provide evidence for oscillatory entrainment phenomena during speech processing, by showing that a rhythmic stimulation elicits subsequent rhythmic activity ('resonance'), either in the behavioural performance or the neural activity, after the stimulation ceases (Cabral-Calderin & Henry, 2022; Henry *et al.*, 2025; Hickok *et al.*, 2015; Kösem *et al.*, 2018; van Bree *et al.*, 2021; Zoefel *et al.*, 2024), or in the absence of rhythmic acoustic fluctuations (Zoefel & VanRullen, 2015). Further evidence for an entrainment account comes from studies showing rhythmic fluctuations in speech perception after using transcranial alternating current stimulation (tACS) (Ten Oever *et al.*, 2016; van Bree *et al.*, 2021; Wilsch *et al.*, 2018), or effects of the pre-stimulus neural phase on perception (ten Oever & Sack, 2015). In contrast, another recent study observed inter-individual differences, where some individuals showed resonance phenomena and others did not (Assaneo *et al.*, 2021), and yet another study showed no evidence for resonance at the group level (Sun *et al.*, 2022). However, the lack of observed resonance is not necessarily evidence for a lack of involvement of oscillatory processes and, reversely, prolonged activity after stimulation can occur in neural populations that do not display self-sustaining oscillator properties (Doelling & Assaneo, 2021). Moreover, a recent review highlighted some inconsistencies and shortfalls of the above findings and approaches (Atanasova *et al.*, 2025). Overall, other approaches might be more fruitful in probing oscillatory activity. For example, Doelling and Assaneo (2021) suggested specifying neural dynamics by choosing a particular dynamical system as a candidate quantitative model to advance the ongoing debate whether oscillations are involved in speech processing or not.

According to the asymmetric sampling in time approach (Oderbolz *et al.*, 2025; Poeppel, 2003), faster (gamma) brain rhythms in auditory cortex are also relevant for speech processing. Gamma brain rhythms in the left hemisphere are thought to be involved in the processing of phonemic information, while slower delta/theta brain rhythms in the right hemisphere are considered for syllable segmentation and processing of speech prosody. However, hemispheric lateralisation can also be affected by other processes, resulting in a rather heterogeneous view (Assaneo *et al.*, 2019a; Flinker *et al.*, 2019; Giroud *et al.*, 2020). While asymmetric sampling in time suggests that sampling through phase-locking to the acoustics occurs for delta/theta as well as gamma brain rhythms (see also, (Giroud *et al.*, 2024), other accounts suggest a different underlying mechanism for gamma (Giraud & Poeppel, 2012; Shamir *et al.*, 2009). For example, decoding of phonemic information from speech acoustics (ranging from slow, 12 Hz, to fast, 50 Hz) are related to theta-gamma (or -alpha/beta) phase-amplitude coupling (Giraud & Poeppel, 2012; Hovsepyan *et al.*, 2020; Hyafil *et al.*, 2015a; Marchesotti *et al.*, 2020; Pefkou *et al.*, 2017), while phonemic decoding and theta-gamma coupling is rather under-explored.

Further to the putative entrainment at the syllabic and phonemic levels, recent research investigated the role of brain rhythms at several other levels of the linguistic processing. At the phrase and sentence level, a prominent study by Ding and colleagues (2016) postulated that the auditory cortex entrains to these units in the absence of acoustic cues for segmentation, which would clearly indicate that an oscillatory mechanism is involved in linguistic processing. However, there are alternative explanations for this finding (Frank & Christiansen, 2018; Frank & Yang, 2018), which could be an artefact of the used non-naturalistic stimulus material (also referred to as 'toy language'; (Kazanina & Tavano, 2023). The role of brain rhythms in higher level linguistic processing are discussed in the language processing section below. Importantly, additionally to the quasi-rhythmic acoustic cues at the syllabic scale, speech contains prominent rhythmic fluctuations related to prosody, which can provide cues for brain rhythm entrainment (see next section on prosody tracking).

Another relevant question is whether and how potential entrainment at the syllabic and phonemic level is modulated by top-down effects. Top-down effects of higher-level linguistic processing on acoustic speech tracking in auditory cortex were suggested in studies that found slow-frequency tracking of speech acoustics to be enhanced for naturalistic speech vs. backwards speech, attended vs. unattended or unintelligible vs. noise-vocoded speech (Park *et al.*, 2015; Peelle *et al.*, 2013; Rimmele *et al.*, 2015; Zion Golumbic *et al.*, 2013). As these effects were not observed in other paradigms that controlled for acoustic differences across conditions, others argued that acoustic speech tracking in the auditory cortex is not modulated by linguistic top-down effects, whereas speech intelligibility affected other processing levels such as semantic integration (Gillis *et al.*, 2023; Howard & Poeppel, 2010). Evidence that acoustic speech tracking is not modulated by speech intelligibility was seen for fast speech, where only beta brain rhythms were related to intelligibility (Pefkou *et al.*, 2017); however, see: (Doelling *et al.*, 2014). In contrast, isochronous speech (frequency-tagging) paradigms that compared listening to native and foreign speech, or investigated artificial word learning, provided evidence for top-down modulations (Buiatti *et al.*, 2009; Pinto *et al.*, 2022; Rimmele *et al.*, 2023). Further studies also showed that speech rate in specific contexts can affect speech segmentation and acoustic speech tracking, resulting in different speech percepts (Dilley & Pitt, 2010; Kösem *et al.*, 2018). In summary, whether and how acoustic entrainment at the syllabic and phonemic level is modulated by top-down effects remains unclear.

In sharp functional differentiation from low-frequency speech-tracking, event- or goal-related changes of neural alpha oscillatory power have been implicated in many perceptual and cognitive operations (Clayton *et al.*, 2018). This change in alpha power is thought to reflect the attentional load during speech processing (also referred to as listening effort): While alpha power decreases when target speech becomes more intelligible and thus less effortful to process Obleser (Obleser & Weisz, 2012), alpha power increases when distracting speech becomes more intelligible and thus harder to ignore (Wöstmann *et al.*, 2017).

Recent research has shown that slow neural phase also tracks (or rather, allows statistical recovery of) linguistic phenomena (lexical processing: (Brodbeck *et al.*, 2018); semantic dissimilarity: (Broderick *et al.*, 2018); surprisal: (Weissbart *et al.*, 2020). Thus, rigorous control by modelling of purely acoustic tracking phenomena alongside linguistic phenomena in the statistical analysis is a necessary precondition, whenever a more speech- or language-specific conclusion is intended.

*Prosody tracking*

Speech prosody comprises the supra-segmental features of speech such as tone (the rise and fall of pitch), stress (e.g., combined pitch, length, and loudness), and rhythm, conceptualised as the semi-periodic recurrence of sounds (e.g., syllables) over time (e.g., (Paulmann, 2016). These speech features are linked to specific oscillatory activity, with activity (i) beyond 8 Hz specifying intrinsic sound characteristics, (ii) around 2-4 Hz referring to theta rate of syllables, and (iii) slower modulations (below 2 Hz) that indicate the melodic pitch contour of speech or phrasal structure (Chalas *et al.*, 2024; Kotz *et al.*, 2018). Thus, while speech prosody can mark linguistic (grammatical) contrasts and boundaries, there is still uncertainty which oscillatory activities are specific for segmental and supra-segmental speech features given their comparable profiles. In addition, prosody can also convey a speaker's emotional state, attitude, and person-specific characteristics (e.g., sex, age, dominance, attractiveness) (Kotz, 2022; Pell & Kotz, 2021), and in communication speech prosody is multimodal (e.g., combining vocal, facial, gestural, body movements (e.g., (Biau *et al.*, 2022; Chandrasekaran *et al.*, 2009; Giraud & Poeppel, 2012).

Biau and colleagues (Biau *et al.*, 2022) reported an increase in energy fluctuations in the delta (prosodic modulations) and theta range (syllable rate) in multimodal speech devoid of semantic content. These fluctuations increased when temporal asynchrony between auditory and visual prosody increased, indicating that delta fluctuations in particular respond to temporal expectations in multisensory integration in speech (see also (Lamekina *et al.*, 2024). In a review on oscillatory activity in multimodal (emotion) processing, it was proposed that theta synchronisation (here, enhanced theta power) mediates the integration of multimodal stimuli, while alpha synchronisation indicates the inhibition in brain areas resolving uncertainties about stimulus quality (Symons *et al.*, 2016). Gross and colleagues (2013) reported that quasi-rhythmic speech features aligned with slower delta oscillations in the auditory cortex while listening to forward and backward presentations of stories. Together, this evidence suggests that slower delta and faster theta oscillations interface when looking at specific prosodic features over time (e.g., integrating syllable stress and melodic pitch contour), while the successful temporal integration of multimodal prosodies extends from auditory to motor brain areas. Alpha modulations in response to salient (emotional) multimodal prosodies might be a specific event-related response of attention in continuous speech (e.g., stories).

In summary, speech prosody is closely linked to distinct oscillatory brain activity, with delta and theta oscillations playing crucial roles in processing prosodic features and their multimodal integration. This interplay of oscillatory dynamics underscores the complex neural mechanisms underlying the perception and processing of speech prosody, both within and across sensory modalities and in its interface with linguistic structure.

*Involvement of beta oscillations in speech prediction*

The predictive coding framework posits that descending signals convey expectations while ascending signals transmit prediction errors (see section 1.2.c; (Clark, 2013; Rao & Ballard, 1999). Within this architecture, beta oscillations have been proposed to support the top-down propagation of internal models, coordinating predictive processing across hierarchical and temporal dimensions (Arnal & Giraud, 2012; Bastos *et al.*, 2012; Engel & Fries, 2010). Applied to speech, the role of beta becomes particularly interesting given the speech signal's rapid dynamics and multi-level structure. The brain must continuously predict at phonemic, articulatory, syntactic, and semantic levels. Here, we synthesise evidence indicating that beta rhythms do not merely reflect predictions but rather contribute to predictive processing by acting as a temporal coordination scaffold, structuring internal dynamics to align multi-level inferences with unfolding sensory input.

Beta activity has been reported during tasks requiring listeners to engage prior knowledge and anticipate upcoming input. At lower levels, beta in secondary auditory cortex appears to modulate gamma oscillations in primary auditory cortex (Fontolan *et al.*, 2014). Beta has also been linked to phonemic disambiguation, potentially via articulatory priors (Bidelman, 2015), and its recruitment correlates with top-down projections from the left inferior frontal cortex to auditory areas (Alho *et al.*, 2014; Bouton *et al.*, 2018). At higher levels of representation, beta oscillations have been implicated in the maintenance and updating of semantic and syntactic expectations. Beta power tends to increase in contexts of strong lexical or structural predictability (Lewis *et al.*, 2016; Shahin *et al.*, 2009), potentially reflecting a form of anticipatory encoding. For example, beta activity scales with semantic predictability, increasing when upcoming words are more expected (Weissbart & Martin, 2024). Similarly, a pre-activation in the beta band is observed at syntactic integration points, interpreted as reflecting preparatory grammatical structuring (Lewis *et al.*, 2016; Segaert *et al.*, 2018).

While these patterns suggest a functional role for beta in stabilising higher-order predictions, its precise computational contribution remains debated. Some studies show beta suppression under low predictability, while others fail to find consistent effects, raising the possibility that its recruitment depends on specific task demands or the stability of internal context models. Rather than directly encoding predictive content, a compelling alternative view is that beta oscillations provide a dynamic scaffold for aligning distributed neural populations, thus ensuring the precise convergence of predictions generated at different levels with the unfolding sensory input.

This perspective aligns with the idea that speech perception critically depends on the precise temporal orchestration of neural activity across cortical hierarchies, involving long-range interactions, particularly in fronto-parietal and fronto-temporal networks (Engel & Fries, 2010; Kopell *et al.*, 2000; Spitzer & Haegens, 2017; Uhlhaas *et al.*, 2010). One proposed mechanism for this coordination is cross-frequency coupling (section 1.3), where beta rhythms might align with slower (delta, theta) or faster (gamma) oscillations to precisely phase-lock excitability across different regions, thereby regulating the flow of information (Hipp *et al.*, 2011). For example, beta-phase coupling to phrasal-scale rhythms (~0.6–1.3 Hz) in motor cortex has been observed during intelligible speech comprehension (Keitel *et al.*, 2018). In contrast, when speech is highly compressed to the point of being unintelligible, beta power diminishes significantly (Pefkou *et al.*, 2017), suggesting that beta-dependent coordination requires coherent temporal structure in the input. These observations are compatible with the view that beta rhythms do not carry predictive content per se but rather support the temporal structuring of predictive operations. Their presence may be necessary when internal models are well-aligned with sensory dynamics, but not always sufficient, as their absence does not invariably imply comprehension failure. This suggests a role in maintaining the stability and precision of internal models during active prediction, serving as a gating or aligning signal rather than a direct carrier of information.

The critical challenge ahead is to move beyond correlational observations and definitively unravel what beta oscillations truly compute. While beta activity reliably emerges under stable temporal and contextual structures, its involvement spans diverse cognitive domains, from motor preparation and working memory to decision-making. This raises a fundamental question: is beta's role in language a domain-specific mechanism for prediction, or a more general cortical principle for maintaining and coordinating structured internal states across cognitive functions? Addressing this ambiguity demands innovative causal paradigms. Future research must employ targeted neurostimulation approaches (e.g., tACS or TMS at specific beta frequencies) within naturalistic speech settings to directly test beta's impact on comprehension. Concurrently, biophysically informed computational models are essential to mechanistically link oscillatory dynamics to hierarchical inference, detailing how beta precisely modulates information flow and error signalling (Hovsepyan *et al.*, 2023). Ultimately, beta may not offer a singular solution to linguistic prediction, but rather a foundational element within the brain's overarching system of temporally organized neural computation.

*Language processing*

While speech – the sound that transfers the linguistic message – can be described by many acoustic features in both time and frequency domain, these features do not themselves contain the linguistic units and structures that determine meaning. We have seen above that oscillations have been proposed to provide a mechanism to 'break in' to the linguistic signal in the form of sensory entrainment (section 1.1.c). To get to language comprehension, however, the brain must infer, based on its linguistic knowledge, the latent phonemes, words, and the hierarchical structure that determine the meaning of the utterance. Of the mechanisms outlined in Section 1, what we discuss here is most consistent with synchronisation of activity between neural populations, and is, at this moment, agnostic between whether the more precise mechanistic description of that synchrony is communication-by-coherence (or otherwise, see Section 1.1.b). Note that much of the empirical work in this section comes from naturalistic presentation of speech with the behavioural goal of language comprehension. These works thus speak to any concerns that synchronisation and coupling effects interact with more-heavily-controlled experimental designs, where, for example, the presentation of the stimulus or task demands might obscure the neural response as seen in the 'wild' (e.g., (Ding *et al.*, 2016).

Structured, meaningful representations of the input arise through synthesis of endogenously generated representations (i.e., brain states, memory) with sensory representations: perceptual inference (e.g., (Marslen-Wilson & Warren, 1994; Martin, 2016, 2020; Shams & Beierholm, 2010). The previously mentioned peak in power at the phrase- and sentence rate by Ding *et al.* (2016) sparked interest in the role of low frequency oscillations in the computation and representation of this latent linguistic information. Since then, many studies have shown that linguistic representations (syllables, words, and phrases) shape delta-band neural activity (Coopmans *et al.*, 2025; Lo *et al.*, 2022; Meyer *et al.*, 2017; Slaats *et al.*, 2023).

These findings raised the question: does the brain (putatively) entrain not only to speech (theta band), but also to endogenously generated representations (e.g. words and phrases), occurring at a delta timescale? Meyer (2018) proposed that oscillations at different timescales (delta, theta, gamma) perform segmentation and identification of linguistic units at their respective timescales (intonational phrases, syllables, phonemes). This proposal gained a lot of traction (e.g., (Molinaro & Lizarazu, 2018; Prystauka & Lewis, 2019; Rimmele *et al.*, 2018), but it is not without problems. The non-isochronous nature of words, phrases and sentences poses a clear problem for neural entrainment to these structures. In fact, Bai and colleagues (2022) showed that a sentence and a phrase with identical durations led to reorganisation of neural phase responses in the delta and theta bands. In this case, entrainment at the phrase timescale in one item would correspond to entrainment at the sentence time-scale in the other. While the differences in phase clearly show that low-frequency activity is sensitive to these fine-grained linguistic differences, it also highlights that entrainment as a mechanism cannot as such underlie the inference of linguistic units.  So, any direct mapping between linguistic structure

'size' or 'duration' and a particular frequency band seems too simplistic (see also (Meyer *et al.*, 2020).

Beyond findings in the lower frequencies, the gamma band is a frequent locus of effects. For example, gamma activity has been found to be modulated by syntactic structure (Nelson *et al.*, 2017; Peña & Melloni, 2012), semantic congruence (Hald *et al.*, 2006; Rommers *et al.*, 2013) and lexical predictability (L. Wang *et al.*, 2018; Weissbart & Martin, 2024). Besides Meyer (2018), several proposals suggest that gamma activity and low-frequency signals inform each other in language comprehension through cross-frequency coupling. For example, Benítez-Burraco and Murphy (2019) suggest that delta-theta inter-regional phase-amplitude coupling constructs syntactic and semantic features when the phase of delta is synchronised with the amplitude of theta, while beta and gamma sources are phase-amplitude coupled with theta oscillations for syntactic prediction and conceptual binding.

In a different account, Martin (2020) capitalises on delta-gamma phase-locking and the role of synchronisation and desynchronisation. In this proposal, ongoing slow rhythms are coupled with high-frequency activity that reflects inference (the activation of abstract grammatical knowledge in memory). This does not suppose a direct mapping between 'size' of linguistic structure and frequency band. Instead, the architecture is built upon a speech-envelope driven oscillation that is transformed by punctate local field potentials (LFPs; gamma bursts). These gamma bursts reflect perceptual inferences of abstract linguistic features and structures, cued by the oscillators. In this view, the 'entrainment' to higher level structure, and effects in slower frequency bands mentioned above, are driven by internal evoked responses to sensory input: a consequence of the inference process.

Given the challenge of understanding the architecture and mechanisms of the neural populations doing the work from scalp-recorded data, we think that leveraging increasingly powerful forward models of the neural response to naturalistic spoken stimuli offers a promising way forward. The temporal response function (TRF) is a time-resolved multiple linear regression approach that allows for the modelling of neural data as a function of aspects of the stimulus, such as acoustic edges (Tezcan *et al.*, 2023), phonemic information (Di Liberto *et al.*, 2015) but see (Daube *et al.*, 2019), and even higher-level linguistic aspects such as words and phrases (Brodbeck *et al.*, 2018; Gillis *et al.*, 2021; Slaats *et al.*, 2024). This approach can be used to model neural data in the time- and frequency domains, as well as the coupling domain (Weissbart & Martin, 2024). These statistical models of the data, in combination with specifications of theories expressed as computational models, can begin to pick at the class of mechanisms that could give rise to observed data and explain how structured meaning arises in the brain.

Before we can establish the links to oscillatory mechanisms, however, we need a computationally-specified theory of structure building (Blokpoel, 2018; Coopmans *et al.*, 2024; Guest & Martin, 2021; van Rooij & Baggio, 2021). Current proposals that attempt to bridge (computationally-specified) theories of linguistic inference and structure building with oscillatory activity are VS-BIND (Calmus *et al.*, 2020), ROSE (Murphy, 2024), a compositional neural architecture for language based on DORA (Martin & Doumas, 2017; Martin & Doumas, 2019), and STiMCON (Ten Oever & Martin, 2021).

It is exceedingly unlikely that a transparent 1-to-1 mapping between linguistic concepts and neural readouts, nor neural computation, exists. Rather than being vexed by this complex 'joint' of nature, we choose to admire how compelling the problem the brain solves is when it creates languages from vibrations in the air.

### 1.9.b  Speech production and motor involvement in language processing

Speech production is a complex sensorimotor task, requiring the precise temporal coordination of auditory, linguistic, motor, and predictive processes, and engaging a bilaterally distributed cortical fronto-temporal-parietal (Giraud & Poeppel, 2012; Hickok & Poeppel, 2007; Indefrey & Levelt, 2004; Rauschecker & Scott, 2009) and subcortical network, encompassing the supplementary motor cortex, cerebellum, and the basal ganglia (Guenther & Vladusich, 2012; Hickok & Poeppel, 2007; Kotz & Schwartze, 2016).

The anticipation of movement typically involves suppression of oscillatory power in the beta-range (14-30 Hz), while the end of a movement increases synchronisation in the same frequency band (Pfurtscheller & Lopes da Silva, 1999). The preparation of a speech act also modulates alpha (8-13 Hz) and beta oscillations in sensorimotor brain regions (Gehrig *et al.*, 2012; Liljeström *et al.*, 2015; Mersov *et al.*, 2018; Saarinen *et al.*, 2006). The preparatory time-locked dynamics in the alpha-beta frequency bands are described as event-related desynchronisation and associated with the generation of internal models that regulate motor control (Bowers *et al.*, 2013; Engel & Fries, 2010; Gehrig *et al.*, 2012; Klimesch, 2012; Liljeström *et al.*, 2015)). In turn, the metrics of beta-band global coherence in primary motor and auditory regions  are associated with the feedforward transmission of speech motor plans to motor effectors and sensory brain regions, while alpha suppression in auditory cortices might indicate the preparation of auditory cortices  by motor activity  i.e., the anticipation of  incoming auditory feedback (Bowers *et al.*, 2013; Engel & Fries, 2010; Gehrig *et al.*, 2012; Klimesch, 2012; Liljeström *et al.*, 2015). With the end of speech production, an event-related synchronisation shows as an increase in beta-band power (or beta rebound). This process leads to inhibiting the motor system and initiating new motor plans (Engel & Fries, 2010; Pfurtscheller & Lopes da Silva, 1999).

A complementary line of research has focussed on the coupling of auditory and motor cortices in speech production and perception and the possible role of slow frequency brain rhythms (Assaneo *et al.*, 2021; Assaneo *et al.*, 2019b; Kotz & Schwartze, 2016; Morillon *et al.*, 2019; Park *et al.*, 2015). This research revealed individual differences in spontaneously synchronising speech production with speech perception (Assaneo *et al.*, 2019b). The production-perception synchronisation strength was related to the functional connectivity (i.e., tracking of the speech acoustics by theta brain rhythms in a prefrontal and precentral speech-motor region of interest) and the structural connectivity between speech motor areas and auditory cortices. Behavioural research suggests that the auditory-motor coupling strength (as measured through production-perception synchronisation) relates to top-down effects from motor production facilitating speech perception (Assaneo *et al.*, 2021). These findings suggest the involvement of brain rhythms. Importantly, individual differences in auditory-motor coupling strength seem to not only impact basic auditory (Kern *et al.*, 2021) and syllable processing (Assaneo *et al.*, 2021; Assaneo *et al.*, 2019b), but also performance in a word learning task (Assaneo *et al.*, 2019b) as well as the comprehension of continuous speech (Lubinus *et al.*, 2023). Although, these models focus on slow-frequency brain rhythms and thus are likely simplified as they neglect more 'native' brain rhythms of the motor cortex (such as beta and delta), this approach provides relevant insight into how individual differences in auditory-motor coupling relate to brain rhythms involved in speech production and perception.

A one-fits all explanation of frequency modulations in speech production seems too simplistic and several open questions remain. These questions converge on a central inquiry: whether neural oscillations, particularly beta rhythms, accurately capture the intricate interplay of sensorimotor and linguistic processes in speech production and perception. They challenge the traditional separation of production and perception, highlighting their inherent coupling in self-generated speech, and question the feasibility of isolating their neural correlates. Furthermore, they address the critical issue of distinguishing motor-related artifacts from genuine neural signals and assess the adequacy of current speech production models in light of emerging neural data, pushing for a broader exploration of frequency bands beyond beta to fully understand the complexities of sensorimotor and self-monitoring mechanisms.

### 1.9.c   Music and rhythm processing

It may at first appear that the case for a role of oscillations in music perception is more straightforward than in speech, as music, almost by definition, is rhythmic. In fact, natural performances of music contain both intentional and unintentional temporal deviations that significantly drift from isochronicity, leading to more expressive and enjoyable performances (Madison, 2000). Furthermore, hierarchy of rhythms at different time scales combined with this expressiveness can lead to significant blurring of the principal note rates of a musical piece. Factor in multiple performers and the situation becomes even more complicated. Synchronisation to notes (akin to speech tracking) can only be the starting point. In this section, we will discuss the role of oscillations in tracking rhythm, beat, and ultimately groove, the drive in music listening that makes us want to move.

Inspired by the speech domain, several studies (Doelling & Poeppel, 2015; Harding *et al.*, 2019; Keitel *et al.*, 2025; Tierney & Kraus, 2015; Zuk *et al.*, 2021) have now demonstrated that low-frequency neural dynamics (1 - 8 Hz) track the amplitude fluctuations in the acoustic envelope of music. Doelling and colleagues (2019) investigated the neural mechanisms of this tracking using simulations and established that it was more plausibly the result of sensory entrainment rather than evoked responses to each note (see discussion in section Speech processing and entrainment). Furthermore, the synchrony was band-limited, with a lower limit of about 1 Hz, below which nonmusicians showed no synchrony without training (Doelling & Poeppel, 2015). This mechanism of synchrony to the note rate is thought to be related to temporal prediction and attention (Arnal *et al.*, 2015; Lakatos *et al.*, 2013; Large & Jones, 1999), whereby the phase of the synchronous oscillator can be used to track expected rhythms and upcoming notes.

In conjunction with this low-level tracking system, beta oscillations have been associated with a more volitional control of rhythmic tracking. Fujioka and colleagues (2012) found that these oscillations, sourced in motor regions, would cyclically rise at a slope corresponding to the tempo of isochronous beats. More recent literature (Biau & Kotz, 2018; Criscuolo *et al.*, 2023; Fujioka *et al.*, 2015) has shown that this beta mechanism is sensitive to beat and metrical structure more so than low frequency oscillations. These findings and their source in motor areas suggest beta oscillations as a flow of information from external sources inward (Arnal, 2012) to modulate low-frequency tracking based on higher order information via phase-amplitude coupling (Arnal *et al.*, 2015), allowing for the extraction of beat predictions from more complex rhythmic sequences through action simulation. However, given more recent literature on the bursting nature of motor beta oscillations (see section 1.2.a; (Jones, 2016), an intriguing new avenue to explore could identify whether single trial analyses in beat processing reflect sustained or transient beta activity. Furthermore, whether this phase amplitude coupling cannot be better explained as deviations from sinusoidal waveform shape (Cole & Voytek, 2019) has not yet been tested.

While experimental findings have suggested a role for beta in the tracking of complex patterns, computational work has somewhat questioned its utility. Large and colleagues (2015) have shown that effective beat extraction can be achieved through interactions of two arrays of oscillators, spanning the frequency space of typical musical beats (~1-5 Hz). By this work, sensorimotor interactions are manifested not through delta-beta coupling but by the interactions of low-frequency oscillators present in both auditory and motor areas. This proposal is also in keeping with literature showing low-frequency oscillations in motor areas and coupling between auditory and motor regions (Assaneo & Poeppel, 2018; Morillon *et al.*, 2019). As such, whether beta plays a computational role in beat tracking or is instead driven by the low-frequency coupling remains an open question. Causal evidence, likely using stimulation techniques, will be necessary to tease out which mechanisms are critical to rhythm and beat perception.

These interactions of delta and beta oscillations in auditory and motor regions may lead to a cognitive sensation of groove – the pleasurable urge to move – in music (Witek *et al.*, 2014). Recent modelling work has shown that this phenomenon can be explained through the interactions of three sets of coupled oscillators Zalta (Zalta *et al.*, 2024): an 'auditory' set which receives the rhythm as input, a 'motor' set which couples to the auditory set and sends temporal predictions (in line with the previous section), and a third set, which receives contrastive input from both. This third group shows greater activity with increasing ratings of groove to syncopated beats. Experimentally, Zalta and colleagues (2024) showed that the delta-beta coupling system described above can be related directly to this model, showing delta in auditory regions is sensitive to syncopation, whereas beta in motor regions is driven by increased groove, perhaps providing a role for beta as the third set of oscillators and receiving modulatory input from motor and auditory delta.

In iterating the proposed roles of neural oscillations in the processing of musical rhythm, this section has particularly highlighted the role of phase-coupling both within and across frequencies as a mechanism for incorporating temporal predictions between sensory and motor cortices. The role of the motor cortex as a source of prediction has a long standing in cognitive science (Arnal, 2012; Cannon & Patel, 2021; Halle & Stevens, 1962; Schwartze & Kotz, 2013) whereby neurophysiology designed to generate actions can be used to generate and compare stimulus predictions in content and time as well. The work described above demonstrates how coupled oscillators may implement this interaction between motor and sensory systems to support complex temporal processing like rhythm and groove.

Still many open questions remain. First, some evidence suggests that these mechanisms might not only implement temporal predictions in rhythms but also carry predictive information of content, pitch, and melody (Chang *et al.*, 2018; Doelling & Poeppel, 2015). Second, it remains unclear how much of these proposed oscillatory mechanisms would be specific to music. Certainly, the mechanisms are inspired by theories in other fields, including speech, attention, and motor domains. Does music hijack the same circuitry? Or are there dedicated regions/dynamics specific to music processing? We may find that the lower-level processing of individual notes could be shared with other regions whereas higher-order processes like groove become a specific musical process. Lastly, this section has focused on the perceptual aspect of music and rhythm, but the signal is also produced by a performer. How these oscillatory dynamics play out in performer-listener interactions is key to understanding the neurobiological underpinnings of musical development and learning within a piece and across genres.

## Conclusions

In reviewing the literature on oscillatory cognitive neuroscience, we set out to detail our current understanding of how basic electrophysiological mechanisms and functions – which manifest as rhythmic activity – underpin human cognition. In doing so, two things became clear: On one side, drawing such connections possesses large explanatory power and has introduced intriguing new perspectives. Take the example of the perceptual sampling idea (VanRullen, 2016): It goes against our intuition of a continuous sensory environment yet provides a framework in which sensory exploration can be understood as a consequence of cyclic neural processes that allocate and distribute attentional resources. If these connections can be confirmed, and new ones established, that would take us closer to a comprehensive theory of human cognition, formulated on a neurophysiological level. On the other side, it seems that many questions remain and that there are few findings that are widely accepted and not currently under challenge.

That at least some of the evidence remains mixed or unclear will also be rooted in the replication crisis in cognitive neuroscience, which has been discussed elsewhere (Huber *et al.*, 2019; Ioannidis, 2005; Rajtmajer *et al.*, 2022). However, beyond a methodological perspective, the present review should also serve as a guide on important questions that future research will need to address to develop a more robust understanding of the role of brain rhythms in cognition. This will contribute to a unified theoretical framework of brain function, which will in turn allow us to test ever more specific predictions.

A striking observation is that, if the oscillatory functions described in Section 1 are indeed as fundamental as believed, then we would expect these to feature consistently in all domains of cognition discussed here. For example, exerting excitatory/inhibitory influences is considered to be a fundamental function of brain rhythms (Buzsáki, 2006). Consequently, it should play a role in most cognitive processes and therefore feature widely in the respective research. This does not seem to be the case, as indicated by the (missing) arrows in **Figure 1**, which show that the role of phase in excitation/inhibition is only made explicit in sections on perception and attention. Furthermore, it seems that each area uses partially proprietary concepts that are sometimes not explicitly connected to research at a mechanistic neural level. Additionally, some common terms are not used consistently across areas, making it harder to obtain a clear understanding of underlying mechanistic concepts. An example is the term 'synchronisation', which is used for synchronised activity across brain areas and for power changes (e.g. "event-related de-/synchronisation"). While the underlying mechanism may be similar, this is neither established nor clear. This indicates a need for more basic research to clarify underlying neural mechanisms for the observed phenomena and a more cautious, nuanced use of terminology.

In closing, our review recognises the wealth of advances brought about by explaining human cognition through the lens of rhythmic brain activity. We also raise challenges that the field will need to overcome — above all, whether brain rhythms have the causal role we typically assume. Nevertheless, we strongly surmise that brain rhythms will remain a pivotal link, tying cognitive function to neuronal processes, and hope that this review sets out the questions we need to ask to arrive at a deeper understanding.

# Acknowledgements

We thank Julio Hechavarria and Francisco Garcia-Rosales for valuable comments at an early stage of the project.

All authors are members of the Scottish-EU Critical Oscillations Network (SCONe), funded by the Royal Society of Edinburgh (RSE Saltire Facilitation Network Award to C.K., A.K, S.P, and P.S. (Reference Number 1963).

A.K. received support from the Medical Research Council [MR/W02912X/1] and the Royal Society of Edinburgh (RSE Saltire Facilitation Network Award, 1963). C.K. received support from the Royal Society of Edinburgh (RSE Saltire Facilitation Network Award, 1963) and TENOVUS Scotland (T23-43). C.S.Y.B. received support from TENOVUS Scotland (T23-43). S.B. received support from the Fyssen Foundation, the Fondation pour l'Audition (RD-2016-R; FPA IDA11), the Agence Nationale de la Recherche (ANR-21-CE28-0028); this work has benefited from a French government grant managed by the Agence Nationale de la Recherche under the France 2030 program, reference ANR-23-IAHU-0003. N.B. received support from the Deutsche Forschungsgemeinschaft (Grant BU2400/8-1 and BU2400/9-1). A.C. received support from the Nederlandse Organisatie voor Wetenschappelijk Onderzoek (NWO - HNUXR80789). K.B.D. received support from the Fondation pour l'Audition (RD-2020-10); a French government grant managed by the Agence Nationale de la Recherche under the France 2030 program (ANR-23-IAHU-0003), and the Institut Pasteur (G5, Human and Artificial Perception). L.D. received support from the European Research Council (852139). J.G. received support from the DFG (GR 2024/5 -1,GR 2024/11 -1,GR 2024/12 -1). D.S.K. received support from the DFG (KL 3580/1-1), the IMF (KL 1 2 22 01), and the European Research Council (101162169). C.L. received support from the Max Planck Institute for Empirical Aesthetics. A.E.M. received support for the Lise Meitner Research Group "Language and Computation in Neural Systems" from the Max Planck Society, and by the Netherlands Organization for Scientific Research (NWO) VIDI (grant 016.Vidi.188.029) and Aspasia (grant 015.014.013), and from the European Research Council (ERC-2024-COG 101170162). J.M.R. received support from the Max Planck Institute for Empirical Aesthetics. V.R. received support from the Next Generation EU (NGEU) and Ministry of the University and Research (MUR), National Recovery and Research Plan (NRRP) PRIN 2022 (2022H4ZRSN - CUP J53D23008040006, D DN. 104 02.02.2022 and P2022XAKXL - CUPJ53D23017340001, D DN. 1409 14.09.2022), the Ministerio de Ciencia, Innovación y Universidades, Spain (PID2019-111335 GA-100), and the Bial Foundation (033/22). M.R. received support from the Ministerio de Ciencia e Innovación (MICIIN) and the Agencia Estatal de Investigación (AEI) under the Ramón y Cajal program (RYC2019-027538-I/0.13039/ 501100011033), and the Basque Foundation for Science (Ikerbasque). F.S. received support from the Suomen Kulttuurirahasto (postdoctoral research grant 00242647). B.Z. received support from the Agence Nationale de la Recherche (ANR-21-CE37-0002), and the Fondation pour l'Audition (FPA-RD-2021-10). P.S. received support from the Swiss National Science Foundation (10531F_220081). S.A.K. received support from the Bial Foundation (#102/22).

The authors declare no competing financial interests.

# Author contributions

Conceptualisation: A.K., C.K., J.G., S.H., J.M.R., V.R., M.R., G.T., M.W., S.P., P.S., S.A.K.

Writing - Original Draft: A.K., C.K., C.S.Y.B., S.B., N.A.B., A.C., K.B.D., L.D., L.G., J.G., L.-I.K., D.S.K., G.L., C.L., A.E.M., J.O., J.M.R., M.R., F.S., S.S., E.S., L.T., G.T., J.T., D.W., M.W., B.Z., S.P., P.S., S.A.K.

Writing - Review & Editing: A.K., C.K., C.S.Y.B., S.B., N.A.B., A.C., K.B.D., L.D., L.G., J.G., S.H., L.-I.K., D.S.K., G.L., R.A.L., C.L., A.E.M., J.O., J.M.R., V.R., M.R., F.S., S.S., E.S., G.T., J.T., M.W., S.P., P.S., S.A.K.

Funding acquisition: A.K., C.K., S.P.

Visualisation: A.K., C.K.

## Section contributions

| Section | Main Authors |
| --- | --- |
| Introduction | Anne Keitel, Christian Keitel |
| 1 Putative oscillatory mechanisms | Anne Keitel, Christian Keitel, Manuela Ruzzoli |
| 1.1. Mechanisms of oscillatory phase | |
| 1.1.a. Excitation-inhibition cycle in oscillation dynamics | Manuela Ruzzoli, Raquel London, Satu Palva, Luca Tarasi, Vincenzo Romei |
| 1.1.b Synchronisation between neuronal populations | Christian Keitel, Paul Sauseng, Satu Palva |
| 1.1c Sensory entrainment | Anne Keitel, Christian Keitel |
| 1.2 Mechanisms of oscillatory power | |
| 1.2a Macroscopic measures of power and excitation-inhibition balance in oscillation dynamics | Manuela Ruzzoli, Paul Sauseng, Gregor Thut, Christian Keitel, Satu Palva, Anne Keitel |
| 1.2b Gating by inhibition | Gregor Thut |
| 1.2c Predictive coding | Eelke Spaak |
| 1.3 Cross-frequency coupling mechanisms | Felix Siebenhuehner, Satu Palva |
| 1.4 Travelling waves | Laura Dugué, Laetitia Grabot |
| 1.5 Resting-state rhythmic activity | Christina Lubinus, Anne Keitel |
| 1.6 Interaction with other bodily rhythms | |
| 1.6.a Respiration | Daniel Kluger |
| 1.6.b Pupil-linked arousal | Christian Keitel |
| 1.6.c Cardiac rhythms | Antonio Criscuolo, Sonja A. Kotz |
| 1.6.d Gastric rhythms | Joachim Gross |
| 1.6.e Circadian rhythms | Antonio Criscuolo, Jonas Obleser, Sonja A. Kotz |
| 1.6.f Interactions between physiological and brain rhythms | Antonio Criscuolo, Sonja A. Kotz |
| 2 Oscillatory mechanisms and their role in cognition | |
| 2.1 Perception & attention | |
| 2.1.a Visual perception & attention | |
| Visual perception | Christian Keitel, Manuela Ruzzoli, Chris Benwell, Niko Busch, Laura Dugué, Laetitia Grabot, Paul Sauseng, Jelena Trajkovic, Luca Tarasi, Vincenzo Romei |
| Visual attention | Christian Keitel, Manuela Ruzzoli, Chris Benwell, Niko Busch, Laura Dugué, Laetitia Grabot, Gemma Learmonth, Paul Sauseng, Jelena Trajkovic, Luca Tarasi, Vincenzo Romei |
| 2.1.b Auditory perception and attention | |
| Auditory Perception | Benedikt Zoefel |
| Auditory attention | Malte Woestmann, Mohsen Alavash, Jonas Obleser |
| 2.1.c Multisensory perception & attention | Laura-Isabelle Klatt, Christian Keitel, Manuela Ruzzoli |
| 2.2 Memory | |
| 2.2.a Working memory | Laura-Isabelle Klatt, Niko Busch, Paul Sauseng |
| 2.2.b Long-term memory | Danying Wang, Simon Hanslmayr |
| 2.3 Communication | |
| 2.3.a Speech and language processing | |
| Speech processing and entrainment | Johanna Rimmele, Anne Keitel, Sonja A. Kotz, Malte Woestmann, Jonas Obleser |
| Prosody tracking | Sonja A. Kotz |
| Involvement of beta oscillations in speech prediction | Sophie Bouton |
| Language processing | Sophie Slaats, Andrea E. Martin |
| 2.3.b Speech production | Antonio Criscuolo, Johanna Rimmele, Sonja A. Kotz |
| 2.3.c Music and rhythm processing | Keith Doelling, Sonja A. Kotz |
| Conclusion | Anne Keitel, Christian Keitel |

Note that authors are listed in no particular order within each section.